\documentclass[prb, aps,twocolumn,latexsym]{revtex4}

\usepackage{amssymb,color,ulem}
\usepackage{latexsym}
\usepackage{graphicx}
\usepackage{amsmath}
\usepackage{setspace}
\usepackage{fancyhdr}
\usepackage{graphics}
\usepackage{lastpage}
\usepackage{graphics}
\usepackage{wrapfig}
\usepackage{comment}
\usepackage{bbold}
\usepackage{xcolor}

\begin{document}

\title{Giant chirality-induced spin-selectivity of polarons}

\author{Dan Klein}
\affiliation{Department of Condensed Matter Physics, Weizmann Institute of Science, Rehovot 76100, Israel}
\author{Karen Michaeli}
\affiliation{Department of Condensed Matter Physics, Weizmann Institute of Science, Rehovot 76100, Israel}


\begin{abstract}

The chirality-induced spin selectivity (CISS) effect gives rise to strongly spin-dependent transport through many organic molecules and structures.  Its discovery raises fascinating fundamental questions as well as the prospect of possible applications. The basic phenomenology, a strongly asymmetric magnetoresistance despite the absence of magnetism, is now understood to result from the combination of spin-orbit coupling and chiral geometry. However, experimental signatures of electronic helicity were observed at room temperature, i.e., at an energy scale that exceeds the typical spin-orbit coupling in organic systems by several orders of magnitude.  This work shows that a new energy scale for CISS emerges for currents carried by polarons, i.e., in the presence of strong electron-phonon coupling.   In particular, we found that polaron fluctuations play a crucial role in the two manifestations of CISS in transport measurements---the spin-dependent transmission probability through the system and asymmetric magnetoresistance. 
\end{abstract}

\maketitle

\section{Introduction}\label{sec:intro}

	Organic molecules and structures---the building blocks of all organisms---commonly exhibit a well-defined chirality. For example, amino acids, enzymes, and various sugars break inversion symmetry. Louis Pasteur first discovered this property; he observed that shining linearly polarized light on organic molecules in their natural form rotates the polarization axis. Soon after, a connection was made between the chiral structure of the molecules and their optical activity. Recent experiments have found that, in addition to being optically active, chiral organic molecules and structures also exhibit strongly spin-dependent electronic transport~\cite{Ray,Goehler,Xie,Kettner,Abendroth,Fontanesi, review1,El-Naggar2019,Vardeny2019,Sun2020,Beard2020,Yan2020,review2}. Such behavior was completely unexpected, as the molecules are neither magnetic nor do they exhibit strong spin-orbit coupling (SOC). This effect, called chirality-induced spin selectivity (CISS), has been observed in various organic chiral molecules and structures at room temperature and under an applied voltage on the order of $1V$.

	The basic experimental setup for observing CISS is comprised of a layer of molecules adsorbed on a metallic surface~\cite{Ray,Goehler,Sun2020}. Electrons are photo-excited from the metal and pass through the molecular layer, at which point their spin-polarization is measured. Through such experiments, it was discovered that the resulting spin polarization depends strongly on the molecule's chirality: the right-handed (left-handed) molecules preferentially transmit the positive (negative) spin projection.	For example, the intensity of the outgoing electron beam passing through a layer of double-stranded DNA exhibits a ratio of more than four to one between the spin projection along and opposite the propagation direction~\cite{Goehler}. This measurement directly probes the transmission probability through the molecules, $T_{s',s} $, where $s$ ($s'$) is the spin of the incoming (outgoing) electron. The experimental observations thus imply that the probability of an outgoing $s'=\uparrow$ electron is different from that of one with $s'=\downarrow$, for unpolarized incoming electrons, i.e.,	 $ \sum_{s=\downarrow,\uparrow}T_{\uparrow,s} \neq \sum_{s=\downarrow,\uparrow}T_{\downarrow,s}$. This property could readily be accounted for if the molecules had been magnetic. However, there is no other evidence of time-reversal symmetry breaking in these systems.	

The observation of the CISS effect indicates that the symmetry between states of different helicity, i.e., the projection of the electron spin on its velocity, is broken. Theories~\cite{Yeganeh,Medina,Gutierrez,Guo,Eremko,Rai,Guo2,Medina2,Matityahu,Avishai,Michaeli,Per,Ora} suggest that this asymmetry stems from a combination of the curved geometry of the molecule with SOC.
As a consequence of the weak SOC in organic materials, all these models predict spin-polarization of only a few percent, well below the observed values and in a narrow energy window compared to the temperature at which the effect is observed. Finding the mechanism amplifying this relatively weak spin-dependent transport  is one of the most basic questions in the field.

A potential clue to the missing enhancement mechanism lies in measurements of the current through single molecules~\cite{Xie}, superhelical polymer microfibers~\cite{Yan2020}, multiheme electron conduits~\cite{El-Naggar2019} and two-dimensional chiral hybrid organic-inorganic perovskites~\cite{Vardeny2019,Beard2020} connected to a magnetic substrate. The substrate is magnetized perpendicular to its surface, and the current through these systems is measured in a two-terminal geometry. A comparison of the currents for two opposite magnetization directions reveals a non-trivial asymmetry in response to a voltage bias. This property is similar to giant magnetoresistance (GMR) observed when a current flows through a tunneling junction between two ferromagnets~\cite{Fert}. There, switching the magnetic moment of one ferromagnetic layer results in a significant change in the conductance. In the CISS experiments mentioned above, however, the non-magnetic molecule respects time-reversal symmetry. Consequently, the observed asymmetry in the resistance $R$ is forbidden in linear response~\cite{Buttiker}, where Onsager-Casimir relations~\cite{Onsager, Casimir} constrain it to be an even function of the magnetic field $\mathbf{B}$, i.e.,\ $ R\left(\mathbf{B}\right) = R\left(-\mathbf{B}\right)$. By contrast, this constraint does not apply to GMR, since only one of the two magnets is reversed.  A closer inspection of the experimental data reveals that the system is far from equilibrium, where Onsager's relations no longer apply. Even then, a current that is asymmetric in $\mathbf{B}$ can be obtained only in the presence of inelastic effects~\cite{Segal}. Thus, the two-terminal experiments open  a unique window into interaction effects.
The similarity in magnitude between the asymmetric magnetoresistance and the spin-polarization obtained in scattering experiments hints at their common origin, leading to the conclusion that interaction effects are significant in these systems.

Biological systems where CISS is observed are mostly band insulators, and electron-electron interactions should not play a significant role in their transport properties. By contrast, vibrational modes abound in these soft biological systems at room temperature. As a consequence, charge transfer frequently occurs via polarons~\cite{Warshel,NitzanBook,Blumberger}. The polaron motion along the molecule polarizes the environment, resulting in non-trivial charge dynamics. In particular, each polaron consists of a phonon cloud that substantially increases its mass compared to the electronic one. While the polaron band narrows in the presence of phonons, the SOC remains unchanged. Thus, the energy window exhibiting CISS encompasses a larger fraction of band energies. The time evolution of a polaron in a chiral molecule was studied in Refs.~\onlinecite{Diaz1,Diaz2,Zhang2020} for non-interacting polarons, i.e., accounting for the enhancement of the effective mass alone. They found that the polaron spin is polarized during its propagation through the system. While the mean-field effect may explain the spin-polarization in the scattering experiment, it does not provide the strong non-linearity required for obtaining asymmetric magnetoresistance~\cite{VanWees}.

The present work systematically studies CISS of polarons beyond mean-field, where the interactions between the polaron and the accompanying cloud of phonons modify their dynamics. In Ref.~\onlinecite{Du2020} it was shown that the dynamical readjustment of the environment to a charge entering and exiting the chiral molecule, the Franck-Condon factor~\cite{}, can further increase the energy window over which spin selective scattering is observed. Such a boundary effect is inconsistent with the observed increase in spin-polarization with the molecule length~\cite{Goehler,Vardeny2019}. Moreover, the Franck-Condon factor does not induce the non-linear effects required for generating an asymmetric magnetoresistance. Therefore, we consider here energy exchange between the polarons and the phonons as the former move along the molecule. We derive the polaron currents in both the scattering and magnetoresistance setups within the GW approximation~\cite{Hedin,Louie,Gunnarsson}.

Recently, a theoretical study by Fransson~\cite{Fransson} found a strong asymmetry in the magnetoresistance as a result of a spin-dependent coupling between electrons and chiral phonons. Such phonon modes, which naturally appear in chiral molecules, are sensitive to the system's geometrical structure similar to the trivial acoustic mode. The majority of vibrational modes in molecules are, however, more localized. Our work complements the work in Ref.~\onlinecite{Fransson} and shows that \textit{localized modes} also support the CISS effect. Specifically, we consider interactions with featureless optical phonons and obtain a strongly spin-dependent charge transport in both types of experimental setups.

We find that the unique charge dynamics of the polarons gives rise to spin-dependent transport at all energy scales. As the polarons propagate through the molecule, they polarize their environment by emitting and absorbing phonons. Consequently, they explore a large part of the energy space, including the region of states exhibiting strong spin selectivity. In addition to polarizing the current, the strong fluctuations of the polarons during their motion along the molecule give rise to highly non-linear current-voltage relations. Consequently, the two manifestations of CISS in transport are of similar magnitude. Moreover, we obtained robust spin-dependent transport at energies below the conduction band. Since electrons entering the molecule at such energies cannot propagate without absorbing a phonon, interactions play a crucial role in transport.  The chemical potential of most organic molecules exhibiting CISS resides within the band gap. Consequently, the onset of current in the magnetoresistance measurements is due to such non-linearities, supporting strong asymmetry. 

The remainder of the paper is organized as follows. In Sec.~\ref{sec:model}, we outline the model for CISS.  In Sec.~\ref{sec:Polarons} we derive the expressions for the current in the presence of polarons. Secs.~\ref{sec:Results-Scatter} and~\ref{sec:Results-MR} discuss our results for the scattering and magnetoresistance measurements, respectively. We end the paper with a summary and discussion.

\section{Model for CISS in chiral molecules}\label{sec:model}

The experimental observation of CISS across many different molecules suggests that it is not very sensitive to microscopic details. The questions we propose to address are qualitative in nature. Thus, they are best addressed within a theoretical framework that captures the essential properties of the molecules, such as their helical structure, in a minimal fashion and is suitable for numerical and analytic analyses. Microscopically, we model the chiral molecules as helix-shaped chain of length $Len=Na_0$ 
\begin{align}\label{eq:Hamiltonian}
\mathcal{H}_{\text{mol}}&=\mathcal{H}_{0}+\mathcal{H}_{\text{int}}+\mathcal{H}_{\text{envir}}. 
\end{align}
The chemical bonds with neighboring atoms select a preferred direction $\hat \tau_n$ at each site $n$ (tangential to the helix), which we use as quantization axis for the orbital angular momentum, i.e., we introduce the quantum number $\ell$ as the eigenvalue of $\vec L_n\cdot \hat \tau_n$. The helical structure results in an anisotropic environment experienced at each atom that strongly lifts the degeneracy~\cite{Binghai} between states with the same total angular momentum but different $|\ell|$. Therefore, the mixing between such states is negligble, and  the effective single electron Hamiltonian can be written as
\begin{align}\label{eq:Hamiltonian2}
\mathcal{H}_{0}&=\sum_{n,\ell,s}\tilde{t}_{\ell}\left[c_{n,\ell,s}^{\dag}c_{n+1,\ell,s}+c_{n+1,\ell,s}^{\dag}c_{n,\ell,s}\right]\\\nonumber&+\Delta_{\text{SOC}}\sum_{n,\ell,\ell',s,s'}c_{n,\ell,s}^{\dag}\left[\vec{\sigma}_n\cdot\vec{L}_n\right]_{\ell,s;\ell',s'}c_{n,\ell',s'}. 
\end{align}%
The operator $c_{n, \ell,s}^{\dag}$ ($c_{n,\ell,s}$) creates (annihilates) an electron at orbital state $\ell$ on site $n$ inside the chain $1<n<N$. The quantization axis of the electron spin $s$ is along the center of the helix, which we denote as the $z$-direction, and $\vec{\sigma}$ is a vector of the three Pauli matrices. The first term describes the kinetic energy associated with hopping between neighboring atoms along the helix. The second term has the familiar form of atomic SOC. In the present case, due to little orbital mixing, SOC simply favors spin alignment of a state with $\ell\neq0$ in the direction of the chemical bonds,  $\hat \tau_n$.  
For the helix-shaped chain, the SOC is 
\begin{align}\label{eq:SOC}
\Delta_{\text{SOC}}&\vec{\sigma}_n\cdot\vec{L}_n=\\\nonumber
&\Delta_{\text{SOC}}\ell\left[\chi\sin\frac{2\pi{n}}{\tilde{R}}\sigma_{x}-\cos\frac{2\pi{n}}{\tilde{R}}\sigma_{y}-\frac{b}{2\pi{R}}\sigma_{z}\right] \mathbb{1}_{\ell,\ell'}.
\end{align}
Here $b$ and $R$ are the pitch and radius of the helix, $\tilde R =\sqrt{(2\pi R)^2+b^2}$, and $\chi = \pm$ denotes the handedness of the atomic helix. The spin quantization axis lies along the molecular axis. Notice that in this model, only $\ell \neq 0$ bands experience SOC. These states are typically responsible for charge transfer via organic molecules and, thus, also for CISS.  In particular, we focus here on states with $\ell=\pm1$, i.e., on transport through $p_x$ and $p_y$ orbitals. To further simplify our model, we neglect the last term of the SOC in Eq.~\eqref{eq:SOC}, which has a limited influence on the strength of the spin-dependent transport. 
Our Hamiltonian is the low-energy limit of the models used in Refs.~\onlinecite{Michaeli,Per,Ora, Geyer} for demonstrating the CISS effect.

The environment includes atomic vibrations and localized charges that interact with the electrons. Some of these modes are sensitive to the helical structure of the molecule, for example, the acoustic vibrations. The majority of the modes are, however, relatively localized on a small number of atoms and do not hold information about the lattice structure. To keep the environment featureless, we model it as a set of optical phonon modes~\cite{Frohlich}    
\begin{align}
\mathcal{H}_{\text{envir}} &= \sum_q \Omega_{q} \sum_{n} a_{q,n}^{\dagger} a_{q,n}. 
\end{align}
The operator $a_{q,n}^{\dag}$ ($a_{q,n}$) creates (annihilates) a phonon of frequency $\Omega_{q}$ on site $n$. The scattering of electrons by an optical phonon, which is characterized by the coupling constant $M_q$, is diagonal in space and spin~\cite{Frohlich}     
\begin{align}
\mathcal{H}_{\text{int}} &= \sum_{n,q,\ell,s} M_q c_{n,\ell,s}^{\dagger} c_{n,\ell,s} \left(a_{q,n}^{\dagger} + a_{q,n} \right). 
\end{align}

The Hamiltonian $\mathcal{H}_{\text{mol}}$ is the starting point of our derivations. In the following parts, we study the manifestation of the electron-environment coupling on the spin-dependent transport properties of the chiral molecules. We focus on the polaron limit where the charge carriers consist of electrons surrounded by a cloud of phonons. To calculate the transport properties of a molecule, we need to connect each end to an electrode~\cite{Datta}
\begin{align}
	\mathcal{H}_\text{tot}=&\mathcal{H}_{\text{L}}+\mathcal{H}_{\text{mol}}+\mathcal{H}_{\text{R}}\\
	+&\sum_{\ell,s}\left[\gamma_s^{L}c_{1,\ell,s}^{\dag}d_{L,1,\ell,s}\hspace{-0.6mm}+\gamma_s^{R}c_{N,\ell,s}^{\dag}d_{R,1,\ell,s}\hspace{-0.6mm}+\text{h.c}.\right].\nonumber
	\label{total}
\end{align}
The parameter $\gamma_{s}^{\alpha}$  is the coupling of the molecule to lead $\alpha$. The index $s$ allows to implement a magnetic lead where the coupling is spin-dependent. The left and right leads are governed by a uniform nearest-neighbor hopping Hamiltonian, i.e.,
\begin{align}
\mathcal{H}_{\text{L/R}}=&-\zeta\sum_{n=1}^{\infty}
\sum_{ \ell,s}\left[d_{L/R,n+1,\ell,s}^{\dag}d_{L/R,n,\ell,s}\hspace{-0.6mm}+\text{h.c.}\right].
\end{align}
The operator $d_{L/R,n,l,s}^{\dag}$ ($d_{L/R,n,l,s}$) creates (annihilates) an electron in state $\ell,s$ on site $n$ of lead $L/R$.

\section{Strong electron-phonon coupling - polarons in a chiral system}\label{sec:Polarons}

In the strong coupling limit, the electron motion is accompanied by a cloud of phonons. Such combined identities, also known as polarons,  are the natural quasiparticles of the system~\cite{Holstein,Mahan2000}. These quasiparticles are found by applying the Lang-Firsov transformation 
 \begin{align}
 	c_{n}\rightarrow e^Sc_{n}e^{-S}\equiv c_{n}X_{n},
 \end{align}
where 
 \begin{align}
 	S &= \sum_{n,q,s}\frac{M_q}{\Omega_{q}} c_{n,s}^{\dagger} c_{n,s} (a_{q,n}^{\dagger}-a_{q,n}).
 \end{align}
The operator $X_{n}$ ($X_{n}^{\dag}$) annihilates (creates) the phonons' cloud 
\begin{equation}
  	X_{n} = \exp \left\{\sum_{q,s}\frac{M}{\Omega_{q}}  (a_{q,n} - a_{q,n}^{\dagger}) \right\}.
 \end{equation}

Applying the transformation on the Hamiltonian in Eq.~\eqref{eq:Hamiltonian}, $\bar{\cal{H}}_{\text{mol}}=e^{S}\mathcal{H}_{\text{mol}}e^{-S}$, results in \begin{align}\label{PolaronHamiltonian}
&\bar{\mathcal{H}}_{\text{mol}}^{\ell}= - \sum_{n,s,s'} c_{n,\ell,s}^{\dag} \left( \Delta_{\text{SOC}}\vec{L}_n\cdot\vec{\sigma}_n^{s,s'} + U \delta_{s,s'} \right) c_{n,\ell,s'}\\\nonumber
&+\sum_{q}\Omega_{q}a_{q}^{\dagger} a_{q} +\tilde{t}\sum_{s, n} \left[\lambda_{n+1,n} c_{n,\ell,s}^{\dag}c_{n+1,\ell,s}  + \text{h.c.} \right]\\\nonumber
 	 	&+  \tilde{t}\sum_{s, n} \left[\left( X_{n}^{\dagger} X_{n+1} - \lambda_{n+1,n} \right)c_{n,\ell,s}^{\dag}c_{n+1,\ell,s}+ \text{h.c.} \right]. \label{eq:inf_H}
 \end{align}
Here $\lambda_{n,n+1} \equiv \lambda=\langle X_{n+1}^{\dagger} X_{n} \rangle<1$ denotes the average phononic fluctuations in the absence of electrons and $U = \sum_{q}\frac{M_q^2}{\Omega_{q}}$ is the polaron shift. The polaron-phonon interaction terms in the last line describe the phonons' reorganization to the polaron's new location after a hopping event occurred. The amplitude $\lambda_{n,n+1}$ includes only processes where the phonon state is unchanged, and thus, it merely accounts for their modified potential energy. The last term in the Hamiltonian contains adjustments that involve phonon emission and absorption. Notice that a phonon-mediated electron-electron interaction term also exists, but we are mainly interested in transport properties at energies within the bandgap. Therefore, it is reasonable to assume very few electrons are involved in the process. Moreover, properly accounting for the effect of electron-electron interactions on CISS also requires adding Coulomb repulsion~\cite{Fransson2} and is beyond the scope of this work. 

Within mean-field theory, i.e., neglecting the last term in Eq.~\eqref{PolaronHamiltonian}, the sole effect of the polarons is to reduce the hopping amplitude to $t=\tilde{t}\lambda$. Equivalently, the polarons' mass is larger than the one of the electrons, and their band is narrower. The mean-field theory is an extreme case of the large polaron limit. In the next section, we show that band narrowing can enhance the manifestation of CISS in scattering experiments (as demonstrated in Refs.~\onlinecite{Diaz1,Diaz2,Zhang2020}), but it cannot capture the asymmetry in magnetoresistance. The last term in the Hamiltonian facilitates hopping events that are accompanied by phonons' emission and absorption. Below, we show that these fluctuations can explain the large polarization in both scattering and transport experiments.

\subsection{Mean-field theory approximation}\label{sec:MF}

\begin{figure}[t]
       \includegraphics[width=0.45\textwidth]{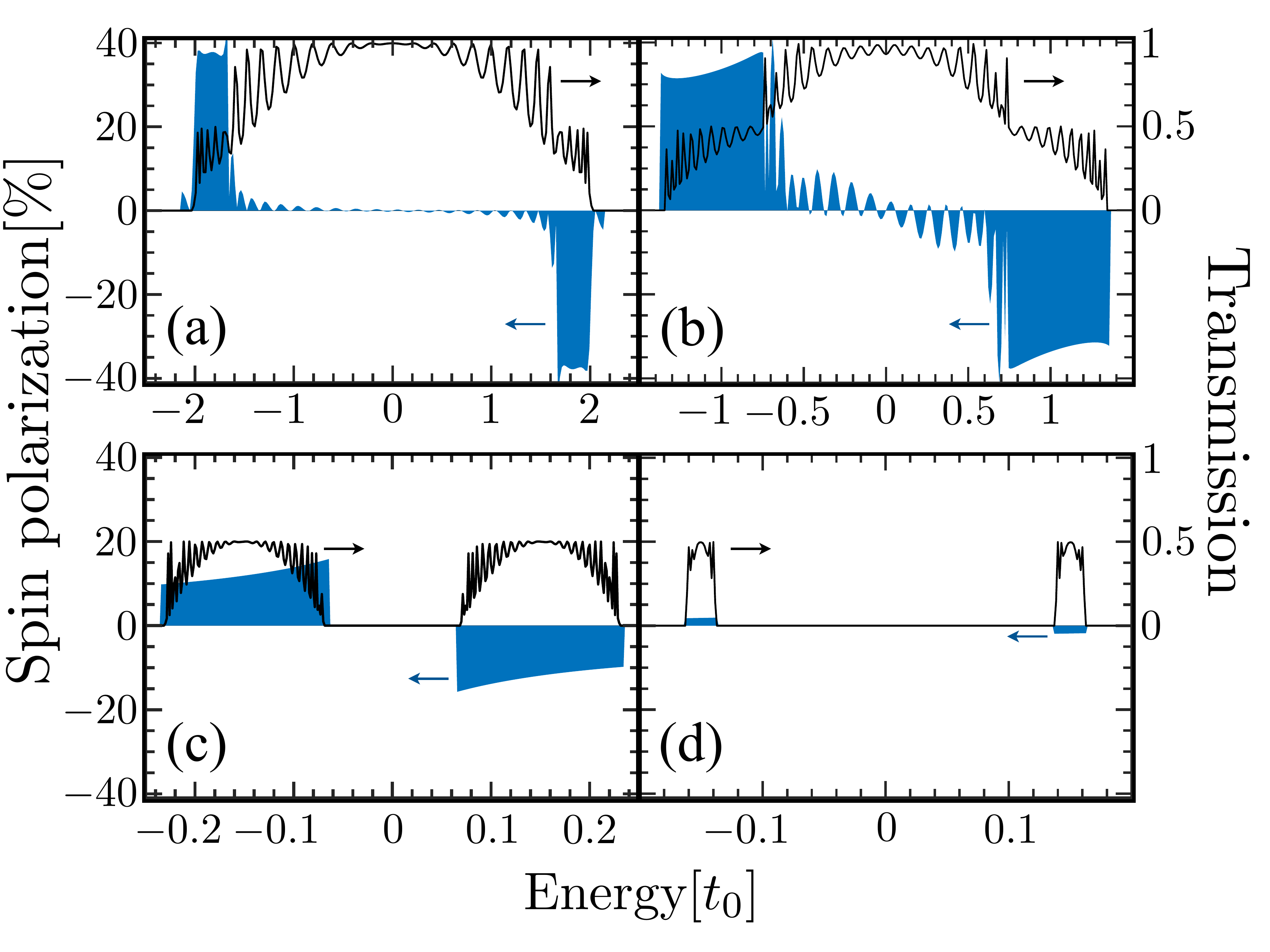}
                \caption{The transmission probability (black) and the corresponding spin-polarization (blue) calculated within mean-field theory as a function of energy for four different temperatures: (a) $T=0.02 t_0$, (b) $T=0.3 t_0$, (c) $T=0.675 t_0$ and  (d) $T=1.05 t_0$. The transmission illustrates the narrowing of the band with increasing temperature. Moreover, as a consequence of SOC, the energy spectrum splits into two subbands  at high enough temperatures $t(T) \ll |\Delta_{\text{SOC}}\ell|$.  At the lowest temperature (panel a) only small fraction of the band $|\Delta_{\text{SOC}}\ell|/t$ exhibit significant spin polarization.  This fraction grows with temperature and eventually the transmission at all energies is spin-dependent (panel b). As $t(T) \ll |\Delta_{\text{SOC}}\ell|$, however, the spin polarization of all states decreases with increasing $T$ (panel c and d). The calculation is done for a single optical mode with $\Omega_0=0.1t_0$ and $M=0.05t_0$, where $t_0$ is the hopping at zero temperature; the SOC amplitude is $\Delta_{\text{SOC}}=0.15t_0$. }
\label{Fig.1}
\end{figure}

Within mean-field theory, the polaronic Hamiltonian~\eqref{PolaronHamiltonian} describes free particles on a chiral lattice with a temperature-dependent hopping amplitude  
\begin{align}
t(T)=\tilde{t}\lambda=\tilde{t} \exp\left\{ -\sum_q\left( \frac{M_q}{\Omega_{q}}\right) ^ 2 \mathrm{coth} \left(\frac{\Omega_{q}}{2 T}\right)\right\}.
\end{align}
The hopping amplitude and, correspondingly, the band's width become smaller as temperature increases. By contrast, the SOC remains independent of temperature.   The strength of the CISS effect is quantified through the spin-polarization  $\mathcal{P}(\varepsilon)=[\mathcal{T}_{\uparrow}(\varepsilon)-\mathcal{T}_{\downarrow}(\varepsilon)]/[\mathcal{T}_{\uparrow}(\varepsilon)+\mathcal{T}_{\downarrow}(\varepsilon)]$, where $\mathcal{T}_{s}(\varepsilon)=\sum_{s'}T_{s',s} (\varepsilon)$ is the transmission probability of a particle with incoming spin $s$. The spin-dependent transmission probability to pass through such a system of non-interacting particles connected to two leads can be straightforwardly calculated~\cite{Datta}.

The spin-polarization in the absence of interactions is determined by a delicate interplay between the spin-dependent hopping amplitude and the SOC (see the analysis in Refs.~\onlinecite{Michaeli,Per,Ora}). To clearly see the role of these two terms in building up the CISS effect, we use the spin-dependent transformation $c_{n, \ell,s}=e^{-i\chi s\pi n/\tilde{R}}{f}_{n, \ell,s}$ on the mean-field Hamiltonian and obtain 
\begin{align}\label{eq:HamiltonianMF}\nonumber
\mathcal{H}_{MF}&= -U\sum_{n,\ell,s} f_{n,\ell,s}^{\dag} f_{n,\ell,s}+\sum_{n,\ell,s}\left[\tau_s f_{n,\ell,s}^{\dag}f_{n+1,\ell,s}+\text{h.c.}\right]\\
&-\Delta_{\text{SOC}}\sum_{n,\ell,s,s'}f_{n,\ell,s}^{\dag}\ell \sigma_y^{s,s'}f_{n,\ell,s'}. 
\end{align}
In this basis, the hopping parameter is spin-dependent, $\tau_{s} = t e^{- i \chi s \pi / \tilde{R}}$. The hopping and SOC terms tend to align the spins in different directions: The former acts as an effective magnetic field along the  $z$-axis that is proportional to the polaron momentum. Consequently, it splits the polaron spectrum according to their helicity ---the projection of the spin on the momentum $h\equiv\hat{k}\cdot\vec{s}=\pm1/2$. This splitting alone, however, does not lead to spin-dependent transport. The SOC is an effective magnetic field in the $\ell\hat{y}$ direction. In the limit, $\Delta_{\text{SOC}}\ll t(T)$, the main effect of the SOC term is to gap states of one helicity while keeping the other intact. As a result, spin-selectivity is obtained at energies within this partial gap. Since the gap is proportional to $\Delta_{\text{SOC}}$, only a small fraction of the band supports CISS. In the opposite limit $\Delta_{\text{SOC}}\gg t \left( T \right)$, the polarons get spin-polarized along $\ell\hat{y}$, thereby lacking a well-defined helicity. No CISS effect can be seen in this limit as different orbital states $\ell=\pm1$ are spin-polarized in opposite directions.  Thus, we expect spin-polarization to depend on temperature non-monotonically. For a detailed discussion of the mean-field Hamiltonian see Refs.~\onlinecite{Michaeli,Per,Ora}.

In Fig.~\ref{Fig.1} we present the spin-polarization $\mathcal{P}(\varepsilon)$ as a function of the energy, for four different temperatures. At low temperature, Fig.~\ref{Fig.1} (a), polarization is seen only at a small fraction of the band. At temperatures where $\Delta_{\text{SOC}}\lesssim t(T)$ the window of energies that shows CISS becomes comparable with the bandwidth, see Fig.~\ref{Fig.1} (b). Finally, as illustrated in Fig.~\ref{Fig.1} (d), the polarization vanishes at a high enough temperature.

\begin{figure}[t]
       \includegraphics[width=0.45\textwidth]{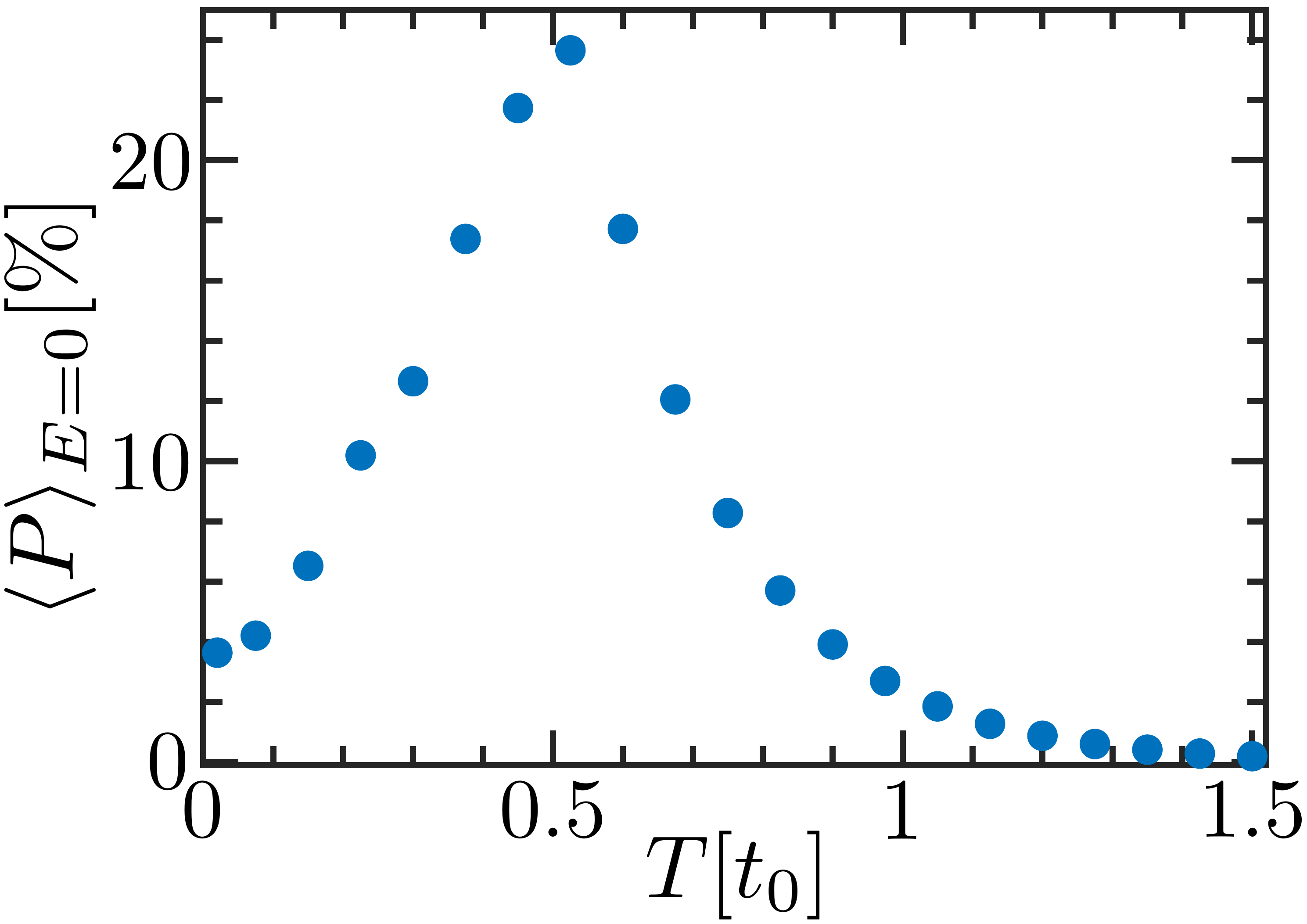}
                \caption{The average mean-field spin-poalrization of the lowest half band as a funciton of temperature. The calculation is performed for the same parameters as Fig.~\ref{Fig.1}.}
\label{Fig.2}
\end{figure}

To quantify the strength of the CISS effect, we calculate the polarization of the average transmission, 
\begin{align}
\langle \mathcal{P}\rangle_{E}=\frac{\sum_{\varepsilon=-\infty}^{E}[\mathcal{T}_{\uparrow}(\varepsilon)-\mathcal{T}_{\downarrow}(\varepsilon)]}{\sum_{\varepsilon=-\infty}^{E}[\mathcal{T}_{\uparrow}(\varepsilon)+\mathcal{T}_{\downarrow}(\varepsilon)]}.    
\end{align}
The average here is taken over energies in the window $\varepsilon<E$, and it corresponds to sending an electron beam with a wide range of energies. For convience, we ignore the constant change in energy inflicted by the polaron shift. The average polarization in the lower half-band $\langle \mathcal{P}\rangle_{E=0}$ as a function of temperature is shown in Fig.~\ref{Fig.2}. The figure clearly illustrates the expected non-monotonous dependence of the CISS effect on temperature. We anticipate, however, that the experiments on organic molecules can only observe an enhancement of the CISS effect with temperature. The SOC of organic structure, which is on the order of a few meV, is much smaller than the coupling between neighboring sites $\sim100$meV. Consequently, the down turn of the polarization occurs well above room temperature when $t(T)$ becomes comparable to $\Delta_{\text{SOC}}$.

One of the main mysteries of the CISS effect is the discrepancy between its strength and robustness in contrast to the small SOC in organic systems. Our results above reproduce the finding of Refs.~\onlinecite{Diaz1,Diaz2,Zhang2020} that the spin-selectivity of the scattering probability is enhanced in the polaron regime at moderate temperatures. The high-temperature regime, though not accessed in experiment, should exhibit opposite behavior.

To complete our discussion of the mean-field theory, we demonstrate its insufficiency in explaining the asymmetry in magnetoresistance measurements.  For this purpose, we calculate the transmission probability in the presence of one magnetic lead, specifically in the left one, $\Gamma_{\uparrow}^{L}\neq\Gamma_{\downarrow}^{L}$. Reversing the magnetization $M$ corresponds to interchanging $\Gamma_{\uparrow}^{L} \leftrightarrow \Gamma_{\downarrow}^{L}$. Within the mean-field theory, the current as a function of voltage is given by the Landauer formula~\cite{Imry}
\begin{align}\label{Landauer}
I(M)=\frac{e}{\hbar}\sum_{s=\uparrow,\downarrow}\int\mathrm{{d\varepsilon}}\left[f_{\text{L}}(\varepsilon)-f_{\text{R}}(\varepsilon)\right]\mathcal{T}_{s}(M,\varepsilon),
\end{align}
where $f_{{j}}(\varepsilon)$ is the distribution function in leads. For measurement of the magnetoresistance, the leads are at equilibrium with chemical potentials $\mu_{\text{L}}$ and  $\mu_{\text{R}}=\mu_{\text{L}}+eV$; $f_{{j}}(\varepsilon)=[e^{(\varepsilon-\mu_{{j}})/T}+1]^{-1}$ is the Fermi-Dirac distribution function of lead $j$. We found that $I(M)=I(-M)$, as expected in the absence of interactions~\cite{VanWees}. Thus, to understand CISS effect on its different manifestations, we must include hopping events accompanied by phonon emission and absorption.

\subsection{Transport properties beyond mean-field}\label{sec:beyond MF}

\begin{figure}[t]
       \includegraphics[width=0.45\textwidth]{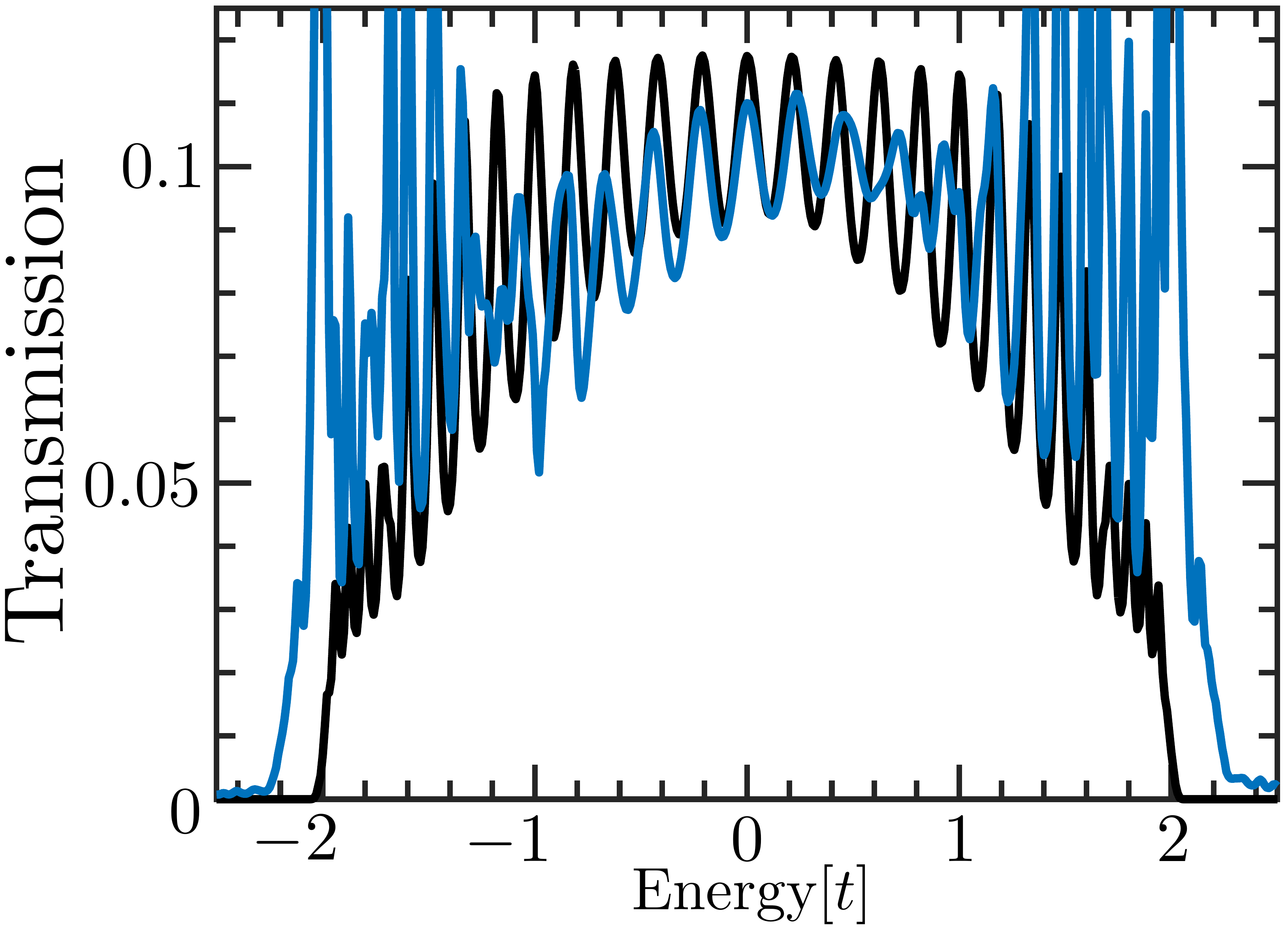}
                \caption{The transmission probability as as function of energy in the presence (blue) and absence (black) of polaron fluctuations. The transmission is found for $T=0.525t$, $\Omega_0=0.35t$ and $M_0=0.09t$.}
\label{Fig.3}
\end{figure}

The molecule's transport properties, those measured in the scattering and in the magnetoresistance experiments, are determined, within mean-field theory, by the transmission probability through the system. This relation to $\mathcal{T}_{s}$ does not hold, however, beyond mean-field. We extract both measurable quantities  from the generalization of the Landauer formula for the current through interacting systems~\cite{Meir}
\begin{align}
&I_{{j}}(M)=\frac{ie}{\hbar}\int{\mathrm{d\varepsilon}}\hspace{-3mm}\sum_{n,n',\ell,\ell',s,s'}\hspace{-3mm}\Gamma_{n',\ell',s';n\ell,s}^{{j}}\\\nonumber
&\times\left\{G_{n,\ell,s;n',\ell,s}^{<}(\varepsilon)[1-f_{{j}}(\varepsilon)]+G_{n,\ell,s;n',\ell,s}^{>}(\varepsilon)f_{{j}}(\varepsilon)\right\}.\label{MW}
\end{align}
The above equation describes the current flowing from the molecule to the lead $j$, and $G_{\mathbf{n}',\mathbf{n}}$ is the interacting Green's function (GF) in Keldysh space~\cite{Rammer}. The superscript $<$ and $>$ denote the lesser and greater components,  $G_{n,\ell,s;n',\ell',s'}^{<}(t)=i\langle c_{n',\ell',s'}^{\dag}(t) c_{n,\ell,s}(0)\rangle$ and $G_{n,\ell,s;n',\ell',s'}^{>}(t)=-i\langle c_{n,\ell,s}(t)c_{n',\ell',s'}^{\dag}(0) \rangle$. The bare  current vertex is $\Gamma_{n,\ell,s;n',\ell',s'}^{{j}}=2\pi \rho_{\ell,s}^{j}\delta_{\ell,\ell'}\delta_{s,s'}\delta_{n',j}\delta_{n,j}$, where $\rho_{\ell,s}^{j}$ denotes the density of states with spin $s$ and orbital $\ell$ in the lead $j$. The GFs are renormalized by both the coupling to the leads and the interactions 
\begin{subequations}
\begin{equation}\label{G<}
G^{<}=i\sum_{j=L,R}f_j(\varepsilon)G^{R}\cdot\Gamma^{j}\cdot G^{A}+G^{R}\cdot\Sigma^{<}\cdot G^{A}.
\end{equation}
\begin{equation}\label{G>}
G^{>}=i\sum_{j=L,R}(f_j-1)G^{R}\cdot\Gamma^{j}\cdot G^{A}+G^{R}\cdot\Sigma^{>}\cdot G^{A}.
\end{equation}
\end{subequations}
The dot product denotes a product of matrices. The self-energy $\Sigma(\varepsilon)$ includes all interaction corrections; the corresponding components of the GF are 
\begin{align}\label{GRA}
\left[G^{R,A}\right]^{-1}&=\varepsilon-\mathcal{H}_{MF}\pm\frac{i}{2}\left[\Gamma^{\text{L}}+\Gamma^{\text{R}}\right]-\Sigma^{R,A}\\\nonumber
&\equiv \left[g^{R,A}\right]^{-1}-\Sigma^{R,A}.
\end{align}

\begin{figure}[t]
       \includegraphics[width=0.45\textwidth]{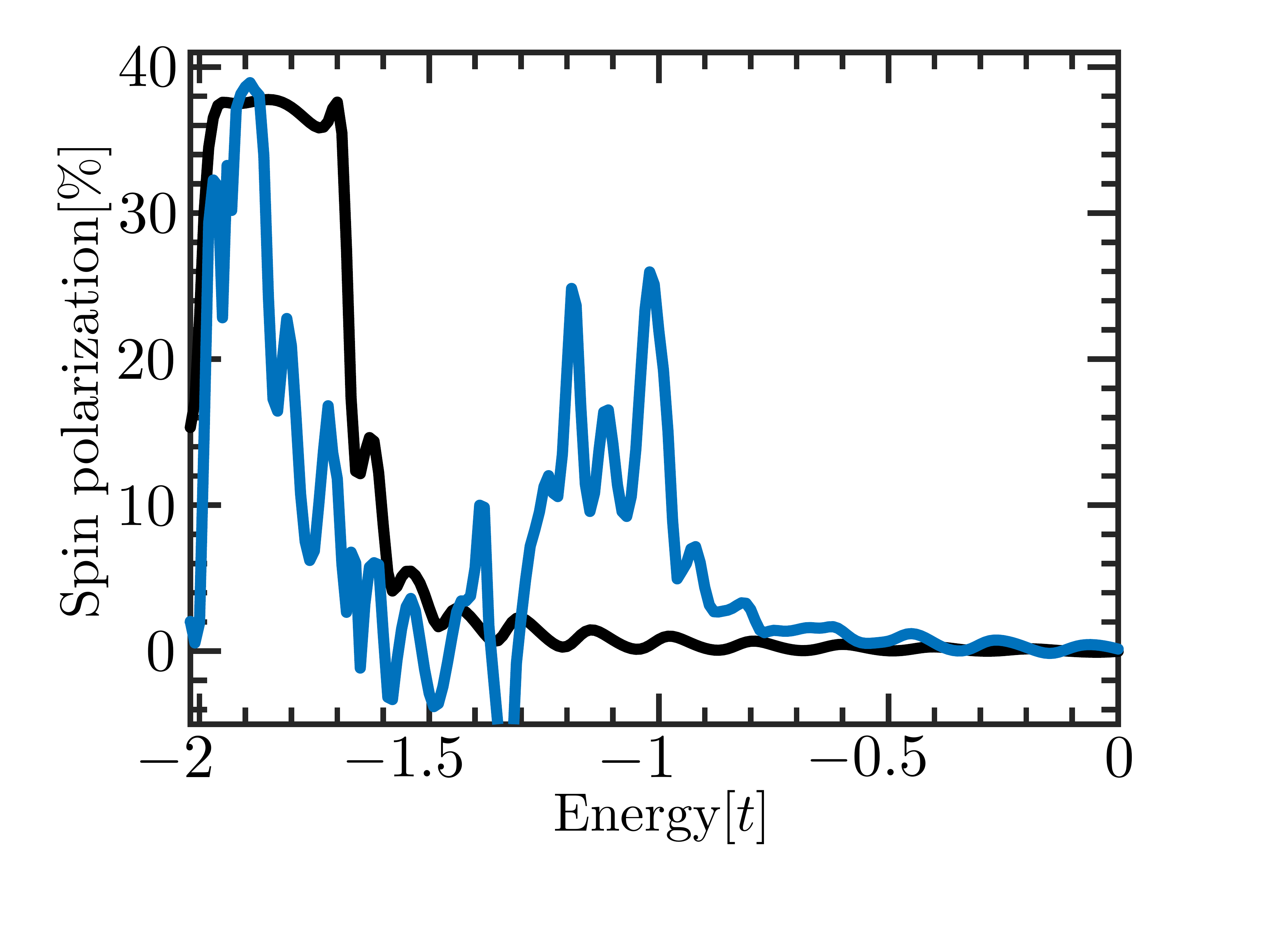}
                \caption{The spin polarization in the lower half band with (blue) and without (black) polaron fluctuations. The polaron fluctuations generate spin polarization at energies where it had been mostly absent. We use the same parameters as in Fig.~\ref{Fig.3}.}
\label{Fig.4}
\end{figure}

To calculate the current, we apply the widely used GW approximation and include only the lowest order corrections to the self-energy~\cite{Klein}. In this non-perturbative approach, the current contains contributions of all orders in the polaron fluctuations. However, we sum over only an infinite subset of corrections in which the self-energy is the simplest.  One of the challenges in calculating the current within the GW approximation is to properly include vertex corrections. For the calculation of the magnetoresistance, we use the expression for the current derived in Ref.~\onlinecite{Klein} for the two-terminal setup. The leads are set to be at equilibrium with two different chemical potentials [see also the discussion below Eq.~\eqref{Landauer}]. In the scattering experiment, on the other hand, the leads are out of equilibrium: (i) The lead where the current is measured is empty, i.e., $f_{L}(\varepsilon)=0$. (ii) The other lead, from which the electrons are injected into the molecule, consists of states at a single energy $\varepsilon_{0}$ and spin $s_{0}$. The corresponding distribution function is  $f_{R}(\varepsilon)\propto\delta(\varepsilon-\varepsilon_0)\delta_{s,s_0}$. Under these conditions, the current in the empty lead is proportional to the transmission probability $\mathcal{T}_{s}(\varepsilon',\varepsilon_0)$. Here the initial and final energies of the electron passing through the system no longer have to be identical, $\varepsilon' \neq \varepsilon_0$. We modify the derivation of Ref.~\onlinecite{Klein} to find a consistent expression for the transmission probability (the equation is shown in Appendix~\ref{App:transmission}).

The main effect of fluctuations is modifying the polaron dynamics. For example, fluctuations generate long-distance polaron hopping that does not exist within mean-field theory (see Appendix~\ref{App:SE}). Specifically for the chiral molecules, the fluctuations introduce spin-dependence into the polaron motion. The latter is already captured by the GW approximation, for which the self-energy is    
\begin{subequations}\label{Self-Energy}
\begin{flalign}\label{Self-Energy<>}
&\Sigma_{n,s;n',s'}^{<,>}(\varepsilon)\\\nonumber
&=\sum_{q,p,p'}\int\frac{\mathrm{d\omega}}{2\pi}D_{n,p;p',n'}^{<,>}(q,\omega)g_{p,s;p',s'}^{<,>}(\varepsilon-\omega).
&&
\end{flalign}
\vspace{-5mm}
\begin{flalign}\label{Self-EnergyRA}\nonumber
\Sigma_{n,s;n',s'}^{R,A}(\varepsilon)&=
\sum_{q,p,p'}\int\frac{\mathrm{d\omega}}{2\pi}\left[D_{n,p;p',n'}^{>}(q,\omega)g_{p,s;p',s'}^{R,A}(\varepsilon-\omega)\right.\\
&\left.+D_{n,p;p',n'}^{R,A}(q,\omega)g_{p,s;p',s'}^{<}(\varepsilon-\omega)\right].
&&
\end{flalign}
\end{subequations}
We emphasize that the coupling to phonons in the polaron Hamiltonian~\ref{PolaronHamiltonian} is spin-diagonal. Nevertheless, the self-energy inherits spin dependence from the polaron propagator. As was shown in Sec.~\ref{sec:MF}, the unique SOC emerging in chiral structures induces spin-selectivity in the electron propagator even in the absence of interactions.   Since all terms are diagonal in $\ell$, we suppress this index below. The propagation of the phonon cloud can be written as $D_{n,p;p',n'}(q,\omega)=V_{n,p} U_{n,p;p',n'}(q,\omega)V_{p',n'}$. The matrix $V_{n,p}=t[\delta_{n,p-1}+\delta_{n,p+1}]$ appears because the polaron fluctuation terms are non-diagonal in the coordinate. The function $U_{n,p;p',n'}= -i \left[ \langle T[ X_{q,n}^{\dag}(t)X_{q,p}(t)X_{q,p'}^{\dag}(0)X_{q,n'}(0)] \rangle-\lambda^2 \right]$ is the (time-ordered) four-point correlation function of the phonon cloud. As discussed in the previous sections, within mean-field theory, a strong spin-dependent polaron motion through the chiral molecules is obtained in a limited energy window, see Fig.~\ref{Fig.1}. The self-energy corrections~\eqref{Self-Energy} extend the spin-dependence to all energies connected to the original narrow range by emission and absorption of phonons.

\begin{figure}[t]
       \includegraphics[width=0.45\textwidth]{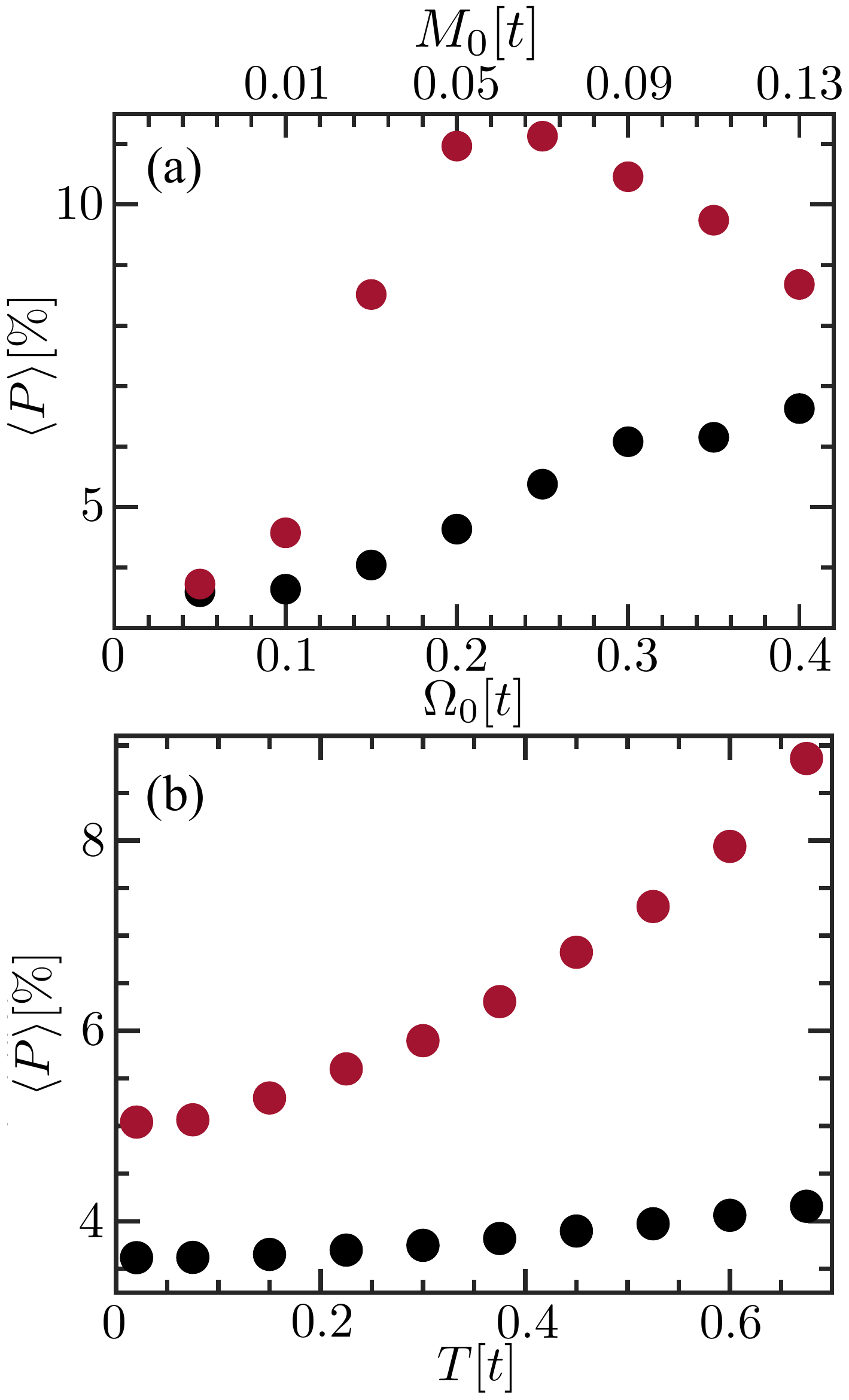}
                \caption{The average spin polarization in the lower half band is presented as a function of (a) frequency at temperatures $T=0.02t$ (red) and  $T=0.525t$ (black). For each frequency, we chose a different coupling strength, which is indicated on the upper $x$-axis. (b) The average spin polarization as a function of temperature in the presence of an optical mode at $\Omega_0=0.15t$ (red) and $\Omega_0=0.35t$ (black).}
\label{Fig.5}
\end{figure}

An additional simplification of the self-energy is obtained by neglecting the renormalization of the phonon modes by the electrons. Such a scenario occurs for a large boson bath, i.e., in the presence of a large number of phonons, as we expect to have in organic molecules.  Consequently, the correlation function of the phonon cloud maintains a simple form \
\begin{align}\label{PhononPropagator}\nonumber
&U_{n,p;n',p'}^{R,A}(q,\omega)=2\lambda^2\sum_{m=1}^{\infty}\left[\frac{1}{\omega-m\Omega_q\pm i\delta}-\frac{1}{\omega+m\Omega_q\pm i\delta}\right]
\\
&\times\left[I_{m}(-y)|\delta_{n,p'}-\delta_{n',p}|+I_{m}(y)|\delta_{n,n'}-\delta_{p,p'}|\right.\\\nonumber
&\left.+I_{m}(-2y)\delta_{n,p'}\delta_{n',p}+I_{m}(2y)\delta_{n,n'}\delta_{p,p'}\right]\sinh\left(\frac{m\Omega_q}{2T}\right).
\end{align}
The index $m$ counts the number of excitations in the cloud, and $I_{m}(y)$ is the modified Bessel function of order $m$ with $y^{-1}=(\omega_q/M_q)^2\sinh(\omega_q/2T)$. The lesser and greater components of the bosonic propagator are $D^{<}(\omega)=N(\omega)[D^{R}(\omega)-D^{A}(\omega)]$ and $D^{>}(\omega)=[1+N(\omega)][D^{R}(\omega)-D^{A}(\omega)]$, where $N(\omega)$ is the Bose-Einstein distribution. See Appendix~\ref{App:SE} for a detailed derivation of the self-energy.

So far, we have provided a general scheme for calculating the charge current through the chiral molecules. We proceed with numerical simulations to obtain results for the polarization in the scattering experiment and the asymmetric magnetoresistance. The parameter space for this problem consists of the molecule's length $Len$, the spin-orbit coupling $\Delta_{\text{SOC}}$, the periodicity of the helix, $\tilde{R}$, the temperature $T$, the electron-phonon coupling $M_q$ and the phonon frequencies $\Omega_{q}$. Our goal is to give a qualitative description of the polaronic CISS effect and not a quantitative one. For simplicity, we assume that band narrowing due to mean-field effects is already substantial at zero temperature and does not change much at the range of temperatures relevant to the experiment.  Thus, in all calculations, we set $\Delta_{SOC}=0.15t$. We assume a single optical mode $\Omega_{0}$ and restrict the sum over the number of phonons in the cloud to be equal unity, i.e., only $m=1$ in Eq.~\eqref{PhononPropagator}. Summing over larger values of $m$ and additional modes clearly enhances the effect.

\begin{figure}[t]
       \includegraphics[width=0.45\textwidth]{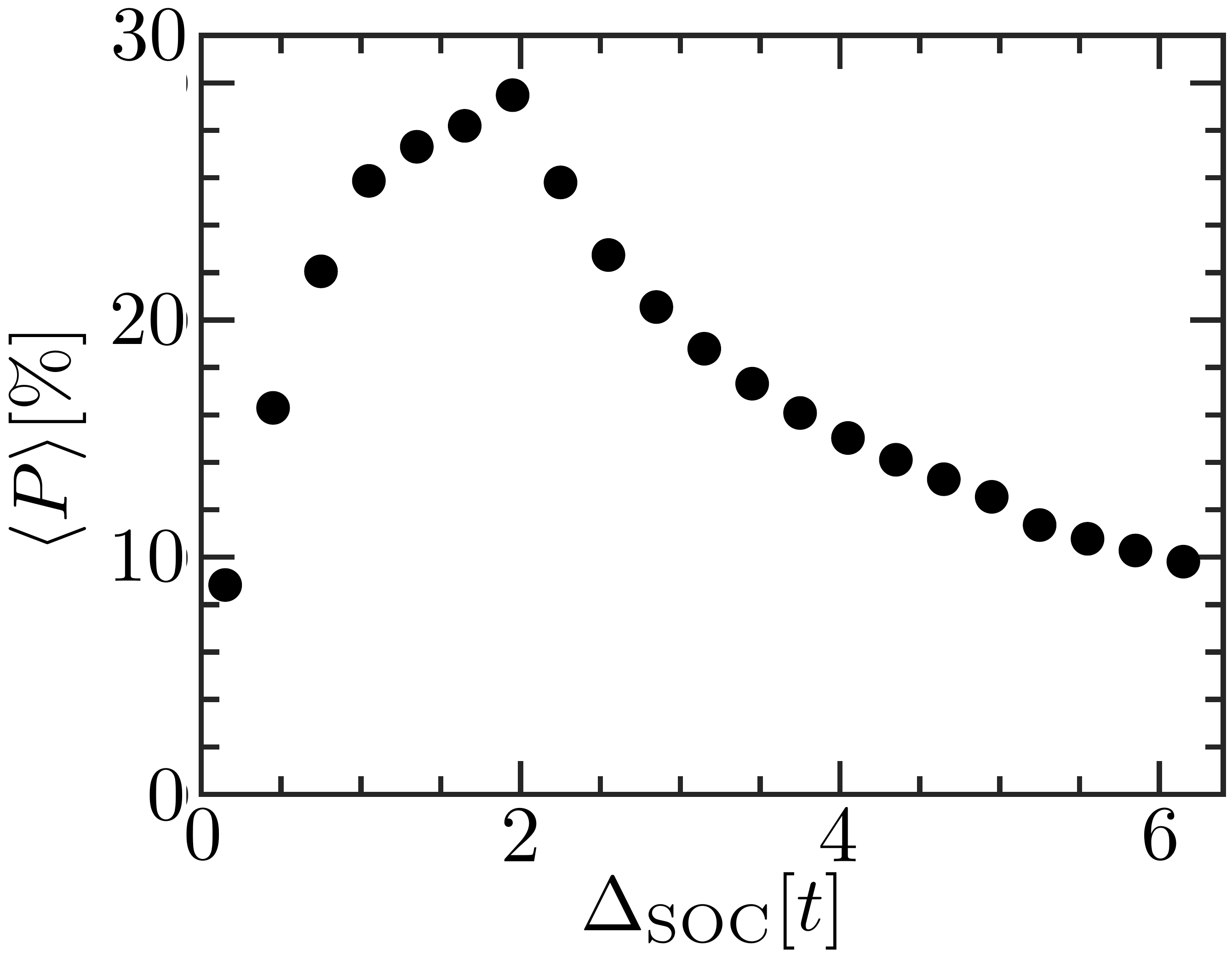}
                \caption{The average spin polarization as a function of $\Delta_{\text{SOC}}$. Similar to the spin polarization as a function of temperature within mean field theory  (\ref{Fig.2}), we find a non-monotonous curve. By contrast, the polarization decreases much more slowly in this case. }
\label{Fig.6}
\end{figure}

\section{Scattering probabilities - results}\label{sec:Results-Scatter}

The scattering and the magnetoresistance experiments bear complementary information on the CISS effect. The former provides the energy-resolved spin-dependent transmission probability and does not require any magnetic component, i.e., time-reversal symmetry is preserved. Magnetoresistance measurements, by contrast, yield only average quantities, but they allow control over parameters such as temperature or the number of molecules. To gain intuition on the origin of CISS, we first performed a thorough calculation of the scattering probabilities through a single chiral molecule. In particular, we fixed the spin $s_0$ and energy $\varepsilon_0$ of the incoming electron beam and found the outgoing current intensity $J_{s_0}$. The latter is proportional to the transmission probability $\mathcal{T}_{s_0}$ and is used to extract the spin-polarization $\mathcal{P}(\varepsilon_0)$. Furthermore, we examine the dependence on the frequencies of the phonon modes $\Omega_q$, the temperature $T$, and the molecule's length $Len$.   

The transmission probability (proportional to the current intensity at the exit) of an incoming beam of unpolarized electrons $J=J_{\uparrow}+J_{\downarrow}$ as a function of energy is shown in Fig.~\ref{Fig.3}.  For the purpose of illustration, we consider a single phonon mode with frequency  $\Omega_0=0.35t$ and set $T=0.525t$. The current is calculated in the presence (blue) and absence (black) of polaron fluctuations. In both cases, the current is substantial at the energy window $-2t<\varepsilon<2t$, where the band forms in the mean-field theory. Polaron fluctuations, however, assist electron hopping and shift the onset of charge transfer to lower energies. Next, we focus on the spin-polarization at energies in the lower half of the original band $-2t<\varepsilon<0$. As explained in the previous sections, our numerical calculations of the current intensity and the corresponding spin-polarization are performed on the simple model Hamiltonian given by Eq.~\eqref{eq:Hamiltonian}. Organic molecules that exhibit CISS typically have a much more complex structure. To focus on the chiral properties of our model and remove other spurious features, we add a small random potential to the Hamiltonian $\delta{H}=\sum_{n,\ell,s}U_nc_{n,\ell,s}^{\dag}c_{n,\ell,s}$, where $U_n$ are randomly drawn from a uniform distribution in the domain $ [-0.1t,0.1t]$.

\begin{figure}[t]
       \includegraphics[width=0.45\textwidth]{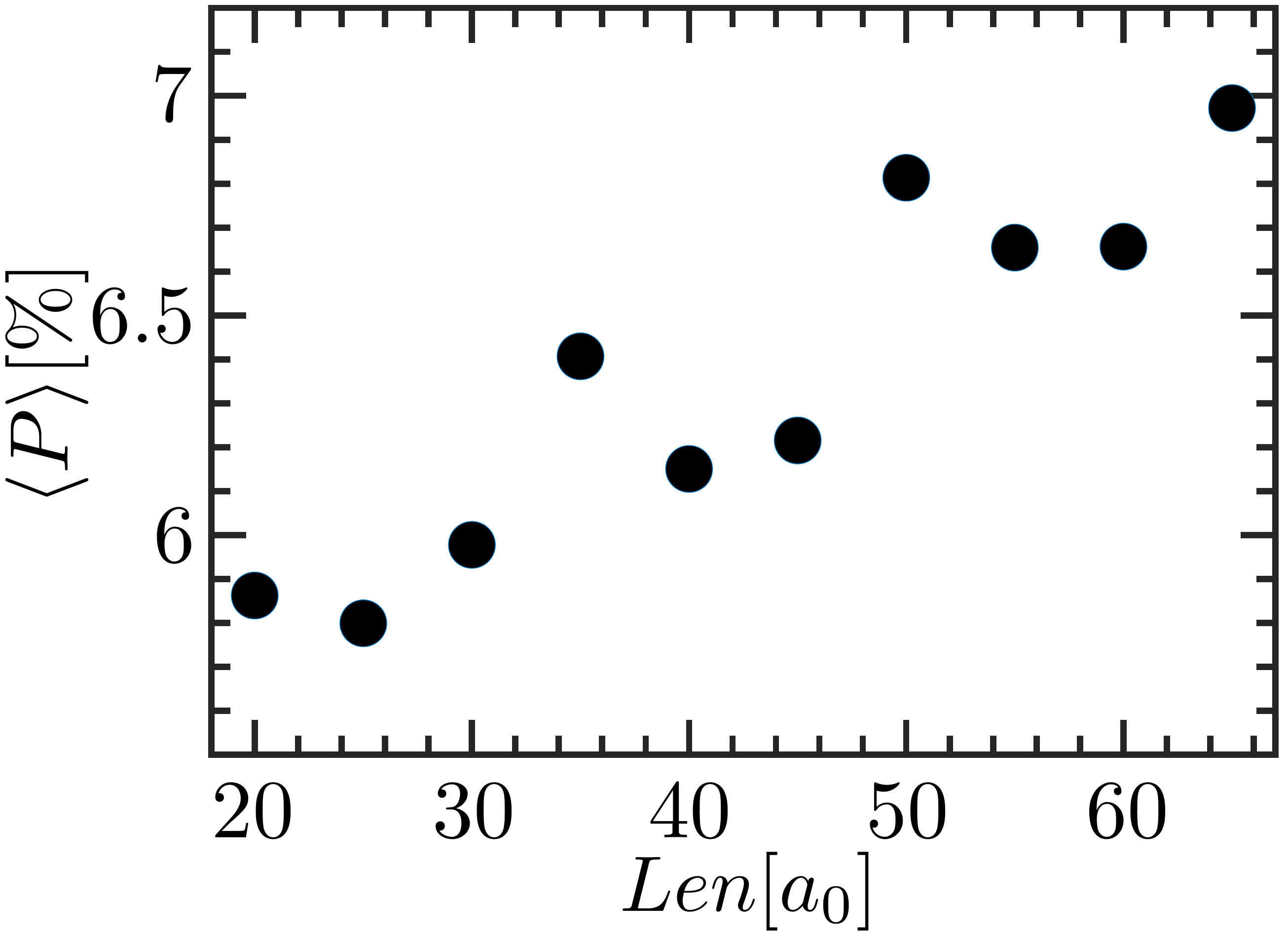}
                \caption{The average spin polarization in the lower half band as a function of length (at $T=0.075t$). }
\label{Fig.7}
\end{figure}

The spin polarization as a function of energy is shown in Fig.~\ref{Fig.4} for the same parameters as in Fig.~\ref{Fig.3}. We see that the polaron fluctuations extend the energies in which $\mathcal{P}$ is substantial. A similar phenomenon occurs in the upper half-band, where the spin polarization is in the opposite direction.  The average polarization in the lower half-band $\langle \mathcal{P}\rangle_{E=0}$ as a function of temperature and frequency is shown in Fig.~\ref{Fig.5}. The polarization grows monotonically with temperature, while its frequency dependence exhibits a maximum at $\Omega\sim2\Delta_{\text{SOC}}$. To understand the above results, we recall that each hop of the polaron is accompanied by reorganization of the environment (phonons). For this adjustment to occur, the polaron emits or absorbs phonons and consequently changes its energy.  Thus, even if the electron enters the molecule at energies where  CISS is weak, it is likely to inherit spin polarization by passing through a chiral state while propagating through the system. Since our calculation of the current within the GW approximation describes processes in which the polaron is scattered several times~\cite{Klein}, the window of energy with large spin-polarization grows by much more than $\Omega_0$. The same exchange of energies with the environment is responsible for lowering the onset of charge transfer well below the bottom of the band. The temperature dependence of the spin polarization is expected as the probabilities of emitting or absorbing phonons increase with $T$.

We recall that we did not account for the temperature dependence of the bandwidth arising from the mean-field corrections. As explained in Sec.~\ref{sec:MF}, the narrowing of the band is equivalent to enhanced SOC. In Fig.~\ref{Fig.6} we present the spin-polarization as a function of $\Delta_{\text{SOC}}$. Similar to the mean-field result (see Figs.~\ref{Fig.1} and~\ref{Fig.2}), the spin polarization decreases as $\Delta_{\text{SOC}}$ becomes larger than $t$. As a consequence of polaron fluctuations, however,  the spin polarization decays more slowly with  $\Delta_{\text{SOC}}$.  Consequently, we expect the downturn of CISS with temperature to be inaccessible for most organic materials. One important step in applying our theory to real molecules would be to find the temperature dependence of the bandwidth within the mean-field theory.

\begin{figure}[t]
       \includegraphics[width=0.45\textwidth]{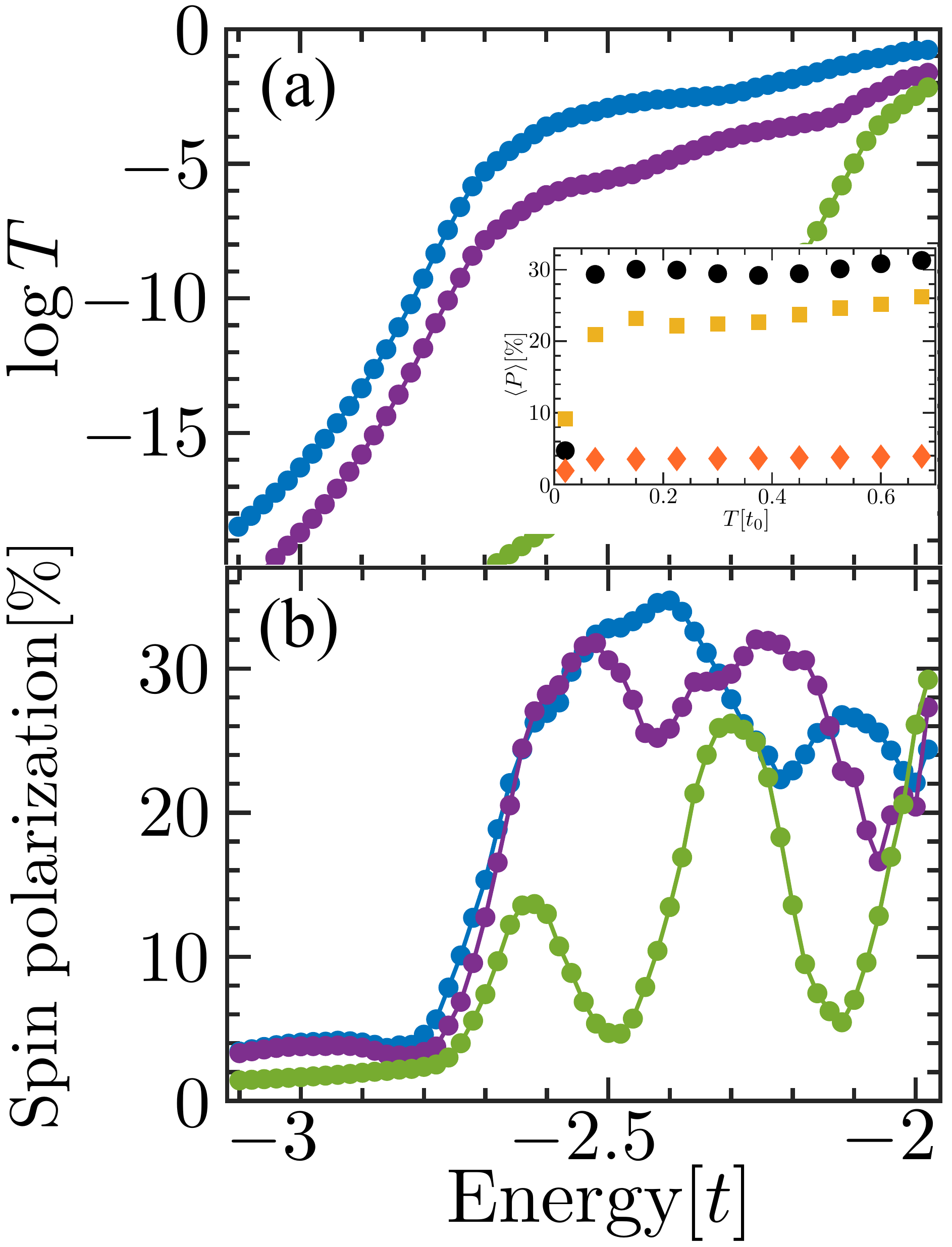}
                \caption{The transmission through the molecule (a) and the corresponding spin polarization (b) at energies below the band for three different temperatures: $T=0.02t$ (green), $T=0.15t$ (purple) and $T=0.675t$ (blue). The transmission grows significantly with temperature, while the polarization saturates to high value $\sim30\%$ already at $T=0.15t$. The temperature dependence of the spin polarization is given in the inset for three different energies: $E=-2.1t$ (black), $E=-2.5t$ (yellow), and $E=-2.9t$ (orange). }
\label{Fig.8}
\end{figure}

Finally, we note that besides temperature and frequency, the probability of a polaron emitting or absorbing phonons during its motion through the molecule is also influenced by the molecule size.  Since the scattering probability is similar on all sites in the molecular chain, the total number of events grows with length. Consequently, we expect the spin-polarization to increase with $Len$.  Such a change in spin-polarization with length has been seen experimentally~\cite{Goehler}, and it is theoretically confirmed by our calculations (see Fig.~\ref{Fig.7}).

So far, we have focused on the polarization at energies inside the original conduction band. However, we have found that polaron fluctuations give rise to significant transmission below the band. The transmission and spin-polarization for energies below the band are shown in Fig.~\ref{Fig.8} for three different temperatures ($T=0.02,0.15,0.675t$, $\Omega_0=0.35t$ and $M=0.09t$). We find that the transmission exponentially decays as the energy of the incoming electron falls below  the bottom of the band. The decay rate reduces with temperature, and as the temperature increases, a significant transmission is obtained at lower energies. Such a phenomenon is frequently observed in measurements of the current-voltage characteristics of organic molecules~\cite{Burin}.  The spin-polarization remains large $\sim25\%$ down to energies with a very low transmission $\mathcal{T}_{s}\ll10^{-5}$ as long as the temperature is not too low. 

A systematic study of $\langle P\rangle_{E}$ at energies below the original band as a function of temperature reveals that it is almost constant, see inset of Fig.~\ref{Fig.8}. A substantial reduction in the spin-polarization is found only at the lowest temperatures. We can understand this result if we recall that charge transfer at energies below the band is governed by phonon absorption, supporting spin-polarization. Thus, $\langle P\rangle_{E}$ remains large as long as the phonon-assisted charge transfer dominates over the direct tunneling.

\section{Magnetoresistance - results}\label{sec:Results-MR}

Measurements of the magnetoresistance are typically performed in a two-terminal setup. In this experiments, chiral molecules are first adsorbed on a magnetic lead. The current through is then measured using a metallic AFM tip as a second electrode~\cite{Xie}, where voltage is applied. Theoretically, we consider two generic leads that are characterized by their density of states. These densities of states enter the GF of an electron inside the molecule through $\Gamma_{\text{L,R}}$, the self-energy corrections due to the leads  [see Eq.~\eqref{GRA}]. Specifically, we implement the magnetic lead via a spin-dependent self-energy, $\Gamma_{\text{L}}^{\sigma_0}>\Gamma_{\text{L}}^{-\sigma_0}$ with $\sigma_0$ being the majority spin. We calculate the current within the GW approximation using the expression derived in Ref.~\onlinecite{Klein}.

In the previous section, we studied the manifestation of CISS in scattering experiments. We found that polaron fluctuations significantly enhance the spin polarization of an electron beam after passing a layer of chiral molecules. In the absence of any magnetic component,  the electronic GF in a system realizing spin-dependent transmission must satisfy
\begin{align}\label{ConditionScatter}
G_{x,\uparrow;x',\uparrow}^{R,A}\neq G_{x,\downarrow;x',\downarrow}^{R,A}.
\end{align}
Without interactions, the inequality given above indicates that $T_{\uparrow,\uparrow}\neq T_{\downarrow,\downarrow}$. Since this condition is satisfied already by Eq.~\eqref{eq:Hamiltonian2}, a small signal of CISS can be observed in scattering experiments even at low temperatures. By contrast, asymmetric magnetoresistance in a two-terminal setup requires a stronger condition 
\begin{align}\label{ConditionMR}
G_{x,s;x',s'}^{R,A}(B)\neq G_{x',-s';x,-s}^{R,A}(-B).
\end{align}
In the absence of interactions, Eq.~\ref{GRA} with $\Sigma^{r,a}=0$ implies that  $G_{x,s;x',s'}^{R,A}(B) = G_{x',-s';x,-s}^{R,A}(-B)$. The structure of the self-energy corrections given by Eq.~\eqref{Self-Energy} implies that Eq.~\eqref{ConditionMR} is satisfied only for systems out of equilibrium. Technically, the distribution functions of the leads entering the self-energy must satisfy $f_{\text{R}}^{\varepsilon}\neq f_{\text{L}}^{\varepsilon}$ (See Appendix~\ref{App:non-eq} for further details). Consequently, observation of CISS in two-terminal transport experiments is made possible solely by interactions and away from linear response~\cite{VanWees}. We verified that our expression for the current is symmetric $I(B)=I(-B)$ at low enough voltages $|\mu_{\text{L}}-\mu_{\text{R}}|\ll\mu_{\text{L}}$ for any choice of $\mu_{\text{L}}$.

\begin{figure}[t]
       \includegraphics[width=0.45\textwidth]{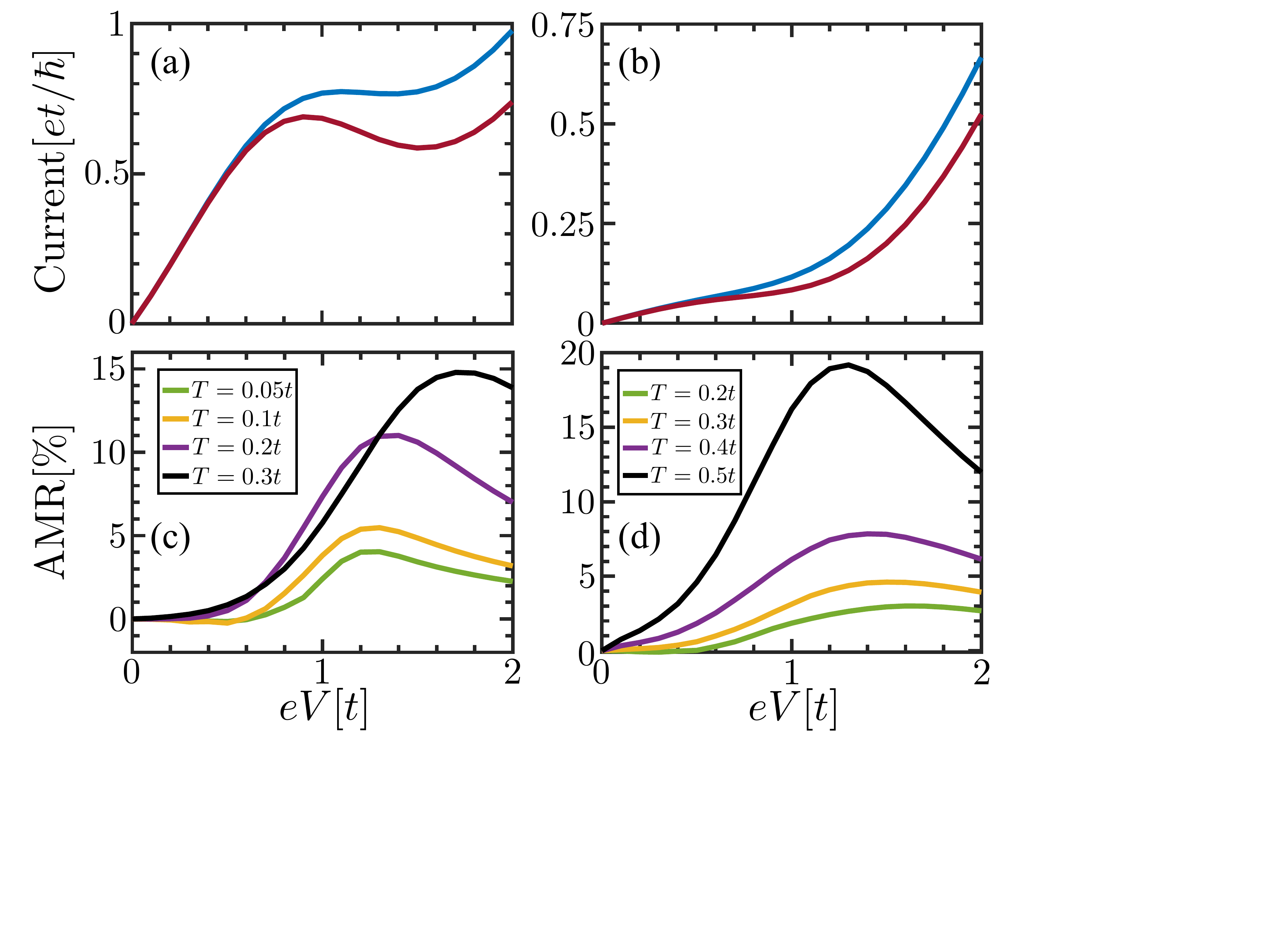}
                \caption{\textbf{Current and asymmetry in magnetoresistance as a function of voltage.} The calculated current of a molecule connected to a magnetic lead with a majority of up (blue) and down (red) spins. The chemical potential of the magnetic lead is $\mu_L=-2.5t$. The chemical potential of the metallic lead is tuned with the voltage $\mu_R=\mu_L+eV$. The electrons are coupled to a phonon mode of frequency (a) $\Omega_0=0.2t$ and (b) $\Omega_0=0.4t$ and the temperature is $T=0.3t$ and $T=0.5t$, respectively. We obtained a large AMR that grows with voltage and temperature, as shown in (c) for $\Omega_0=0.2t$ and (d) $\Omega_0=0.4t$. Coupling to a high-frequency phonon mode gives rise to significant AMR at voltages well below the bottom of the electronic band. }
\label{Fig.9}
\end{figure}

\begin{figure}[t]
       \includegraphics[width=0.45\textwidth]{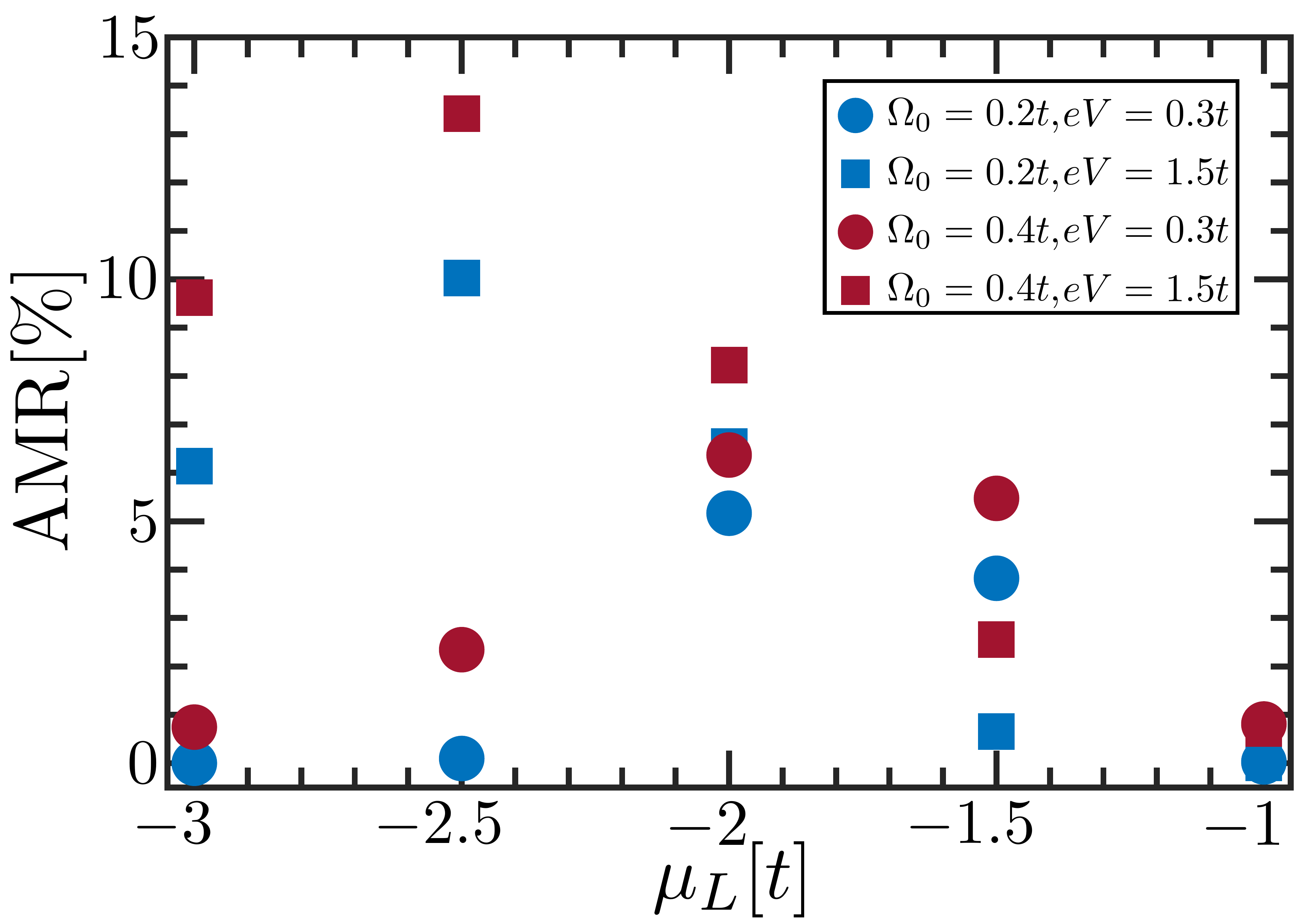}
                \caption{\textbf{The AMR as a function of voltage for different chemical potentials of the magnetic lead $\mu_L$}. The chemical potential of the metallic lead is tuned with the voltage $\mu_R=\mu_L+eV$. The AMR is maximal for $\mu_L$ below, yet close to the bottom of the band, similar to the spin-polarization in the scattering measurement setup.}
\label{Fig.10}
\end{figure}

The discussion above explains that the enhanced spin-dependent transmission does not guarantee a strong asymmetry in magnetoresistance. Nevertheless, our model puts both manifestations of CISS, or the mechanism for both effects, on equal footing. In Figs.~\ref{Fig.9}(a) and~\ref{Fig.9}(b)  we present the calculated current as a function of applied voltage for opposite alignment of the magnetic leads. To mimic the experimental setup, we set the chemical potential of the left lead to be below the bottom of the band $\mu_{\text{L}}=-2.5t$ while the chemical potential of the right lead is tuned by the voltage $\mu_{\text{R}}=\mu_{\text{L}}+eV$. Additional parameters used for the derivation are $\Omega_0=0.4t_0$, $M_0=0.18t$, and $T=0.5t$. We find a strong asymmetry in magnetoresistance $\text{AMR}=[I(B)-I(-B)]/[I(B)+I(-B)]$ on the order of $5\%$ for low currents that grows more with increasing voltage. The AMR is shown as a function of voltage for different temperatures in  Figs.~\ref{Fig.9}(c) and~\ref{Fig.9}(d). As expected, the asymmetry in magnetoresistance grows with temperature. We find that the AMR at $eV<0.5t$, i.e., when both chemical potentials are below the bottom of the band, strongly depends on the temperature and frequency of the phonons. In particular, different phonons give rise to strong asymmetry at different temperatures. In real systems, we expect the phonons' density of state to include a large range of frequencies. Moreover, polaron fluctuations are not limited to the absorption or emission of a single phonon. Thus, we expect to find a significant AMR in a large class of chiral molecules and structures. In Fig.~\ref{Fig.10} we show AMR for different values of  $\mu_{\text{L}}$. This figure supports our conclusion from the previous section that the most significant enhancement of CISS by polaron-phonon interactions occurs at energies below the conduction band. The different calculations of the AMR demonstrate that polaron fluctuations are crucial for our understanding of the mechanism leading to the CISS effect.

\section{Discussion}\label{sec:Discussion}

We have studied CISS in the presence of strong electron-phonon interactions. In this regime, the charge and spin are carried by polarons, whose motion is accompanied by significant polarization of the environment. Our work demonstrates that polaron fluctuations give rise to a strong signal of CISS in both spin-dependent scattering and magnetoresistance measurements. Importantly, it is sufficient that the chiral structure of the molecule enters through the (bare) electronic spectrum while the phonons can be featureless, similar to the optical modes considered here. The polarization of the environment via the emission and absorption of phonons results in the polaron exploring different states in energy space as it moves along the molecule. Thus, charge transfer in such a system is highly non-linear, and its effect on CISS manifests itself in both the spin-dependent scattering probability and the asymmetric magnetoresistance with a similar magnitude. We note that our calculations were performed without optimizing any model parameters.

A key property of our work is that we provide a general framework for calculating transport signatures of CISS within the GW approximation. Our expression for the currents can be applied to more accurate models of chiral molecules to better understand CISS in real systems.  For this purpose, we only need to know the electronic states within the mean-field theory, the phonon spectrum, and the coupling parameters between the electrons and the environment. The physical picture should not change for a complex model as long as it exhibits a well-defined chiral structure, but we hope to be able to fit concrete experimental results in the future.   
 
\section*{Acknowledgments}
This work was supported by Grant No. 2017608 from the United States-Israel Binational Science Foundation (BSF).

\appendix

\section{The polaron self-energy}\label{App:SE}

The GW approximation significantly simplifies the derivation of the polaron GF without neglecting the effect of phonon emission and absorption during its propagation. Here, we present a detailed derivation of the polaron self-energy given by Eqs.~\eqref{Self-Energy} and~\eqref{PhononPropagator}.  We start our derivation in the time domain: 
\begin{align}\label{eq: Sigma_pol}
&\Sigma_{n,s;n',s'}(t_1-t_2)\\\nonumber
&=
	\tilde{t}^2\left[ \Big{\langle} T \left( X_{n}^{\dagger} X_{p} \right)_{t_{1}} \left( X_{p'}^{\dagger} X_{n'} \right)_{t_{2}} \Big{\rangle} - \lambda^{2} \right]   g_{p, s;p',s'} ( t_{1} - t_{2} ).
\end{align}
For $n \neq p$ and $n' \neq p'$, the time ordered phonon four-point correlation function is~\cite{Mahan2000}
\begin{align}
	&-i \Big{\langle} T \left( X_{n}^{\dagger} X_{p} \right)_{t_{1}} \left( X_{p'}^{\dagger} X_{n'} \right)_{t_{2}} \Big{\rangle}\\\nonumber
	& = -i\theta \left( t_{1} - t_{2} \right) e^{\Phi \left( t_{1} - t_{2} \right)} -i \theta \left( t_{2} - t_{1} \right) e^{\Phi \left( t_{2} - t_{1} \right)},
\end{align}
where
\begin{align}\label{Phi}
e^{\Phi \left( t \right)} = \lambda^2 e^{-\left( \delta_{n,n'} + \delta_{p,p'} - \delta_{n,p'} - \delta_{n',p} \right) \phi( t) },
\end{align}
and 
\begin{align}
	\phi \left( t \right) &= \sum_{q}\left( \frac{M_q}{\omega_{q}} \right)^{2} \left( \left( N_{\omega_{q}}+1 \right) e^{-i \omega_{q} t} + N_{\omega_{q}} e^{i \omega_{q} t} \right).
\end{align}

The four-point correlation function of the phonon cloud can be expressed in terms of the modified Bessel function of the first kind $I_{m} \left(y\right)$ using the relations
\begin{equation}
	e^{p \phi \left( t \right) } = \sum_q\sum_{m=-\infty}^{\infty} I_{m} \left( py \right) e^{m\beta \omega_{q} / 2} e^{-i m \omega_{q} t}.
\end{equation}
Here $y^{-1}=(\omega_q/M_q)^2\sinh\left( \omega_{q} /2T\right)$ and the integer $p=\pm1,2$ is determined by the delta functions in Eq.~\eqref{Phi}. 
Finally,  performing Fourier transform of the above identity from time to frequency domain brings us to Eq.~\eqref{PhononPropagator} which is used in the calculation of the self-energy.

\section{The current in the scattering experiment setup}\label{App:transmission}

In the scattering setup, a layer of molecules is adsorbed on a metallic electrode. A beam of photoelectrons with a narrow distribution of energy is extracted from the lead, and the current intensity is measured after the electrons pass through the molecules. The spin of the electron beam can be controlled by choosing an electrode with a large spin-orbit coupling (such as gold) and shining it with circularly polarized light~\cite{Ray,Goehler}. The spin of the electrons exiting the system is often also measured. To derive the expression for the current, we consider a molecule connected to two leads. The first lead, where the current is measured, has all its states empty. The second, where the spin-polarized electron beam enters the molecule, has only one occupied state. Consequently, we use the Meir-Wingreen formula for the current through an interacting finite region with non-equilibrium distribution functions of the left and right leads $f_{\text{L}}(\varepsilon)=0$ and  $f_{\text{R}}(\varepsilon)=\delta_{\varepsilon,\varepsilon_0}\delta_{s,s_0}$. Since we cannot calculate the exact GF of the polarons, we approximate its self-energy to the lowest order in the interaction. This simplification, commonly dubbed the GW approximation, goes beyond a simple perturbation theory. Recently we showed~\cite{Klein} that the corresponding approximate current must also include vertex corrections to be consistent. There, we derived the current for cases where the leads are at equilibrium. We use the expression of Ref.~\onlinecite{Klein} for calculating the magnetoresistance in Sec.~\ref{sec:Results-MR}. Here, we modify the current derivation for the unique distribution functions of the leads in the scattering experiment setup. 

The expression for the scattering current is
\begin{align}\label{CurrentDiagaram}
&J_{\text{L}}(\epsilon)=\mathcal{J}_0\delta_{\varepsilon,\varepsilon_0}\mathcal{I}_a+\frac{\mathcal{J}_0}{2}\delta_{\varepsilon,\varepsilon_0}\int\frac{\text{d}\omega_1}{2\pi}\left[N_{\omega_1}+1\right]\\\nonumber
&\times\left[\mathcal{I}_b-\mathcal{I}_c+\mathcal{I}_d-\mathcal{I}_e+2i\mathcal{I}_f+2i\mathcal{I}_g-2i\mathcal{I}_h\right]\\\nonumber
&
+i\frac{\mathcal{J}_0}{2}\delta_{\varepsilon,\varepsilon_0}\int\frac{\text{d}\omega_1\text{d}\omega_2}{(2\pi)^2}\left[\mathcal{I}_i+\mathcal{I}_j+2i\mathcal{I}_k\right]\left[N_{\omega_1+\omega_2}+1\right]\\\nonumber
&\times
\left[N_{\omega_2}^{\text{ph}}-N_{-\omega_1}^{\text{ph}}\right].
\end{align}
Here, $\mathcal{J}_{0}$ is the intensity of the incoming current, and $\mathcal{I}_{\alpha}$ are the following products of polaron and phonon GFs:
\begin{widetext}
\begin{subequations}
\begin{flalign}
\mathcal{I}_a=\left[\Gamma^{L}\cdot G^{R}\cdot \Gamma^{R}\cdot S\cdot  G^{A}\right]_{n,s;n,s}^{\varepsilon};&&
\end{flalign}
\begin{flalign}
\mathcal{I}_b=&\left[g^R\cdot \Gamma^{R}\cdot S\cdot g^A\cdot \Gamma^{L}\cdot g^R-
g^A\cdot \Gamma^{L}\cdot g^R\cdot \Gamma^{R} \cdot S\cdot g^A
\right]_{n',s',n,s}^{\varepsilon}\left[D^{R}-D^{A}\right]_{n,p;p',n'}^{\omega_1}
\left[G^{R}\cdot\Gamma^{L}\cdot G^{A}\right]_{p,s;p',s'}^{\varepsilon-\omega_1};&&
\end{flalign}
\begin{flalign}
\mathcal{I}_c=&\left[g^R\cdot \Gamma^{R}\cdot S\cdot g^A\cdot \Gamma^{L}\cdot g^R-
g^A\cdot \Gamma^{L}\cdot g^R\cdot \Gamma^{R} \cdot  S \cdot g^A
\right]_{n',s',n,s}^{\varepsilon}\left[D^{R}-D^{A}\right]_{n,p;p',n'}^{\omega_1}
\left[g^{R}\cdot\Gamma^{L}\cdot g^{A}\right]_{p,s;p',s'}^{\varepsilon-\omega_1};&&
\end{flalign}
\begin{flalign}
\mathcal{I}_d=&\left[g^R\cdot \Gamma^{R}\cdot S\cdot g^A\cdot \Gamma^{L}\cdot g^R-
g^A\cdot \Gamma^{L}\cdot g^R\cdot \Gamma^{R}\cdot S\cdot g^A
\right]_{n',s';n,s}^{\varepsilon-\omega_1}\left[D^{R}-D^{A}\right]_{p',n';n,p}^{\omega_1}
\left[G^{R}\cdot\Gamma^{R}\cdot G^{A}\right]_{p,s;p',s'}^{\varepsilon};
&&\end{flalign}
\begin{flalign}
\mathcal{I}_e=&\left[g^R\cdot \Gamma^{R}\cdot S\cdot g^A\cdot \Gamma^{L}\cdot g^R-
g^A\cdot \Gamma^{L}\cdot g^R\cdot \Gamma^{R}\cdot S\cdot g^A
\right]_{n',s';n,s}^{\varepsilon-\omega_1}\left[D^{R}-D^{A}\right]_{p',n';n,p}^{\omega_1}
\left[g^{R}\cdot\Gamma^{R}\cdot g^{A}\right]_{p,s;p',s'}^{\varepsilon};
&&\end{flalign}
\begin{flalign}
\mathcal{I}_f= \left[ g^A\cdot \Gamma^{L}\cdot g^R
\right]_{n',s';n,s}^{\varepsilon-\omega_1}\left[D^{R}-D^{A}\right]_{p',n';n,p}^{\omega_1}
\left[G^{R}\cdot\Gamma^{R}\cdot S \cdot G^{A}\right]_{p,s;p',s'}^{\varepsilon};
&&
\end{flalign}
\begin{flalign}
\mathcal{I}_g=\left[ G^A\cdot \Gamma^{L}\cdot G^R
\right]_{n',s';n,s}^{\varepsilon-\omega_1}\left[D^{R}-D^{A}\right]_{p',n';n,p}^{\omega_1}
\left[g^{R}\cdot\Gamma^{R}\cdot S \cdot g^{A}\right]_{p,s;p',s'}^{\varepsilon};
&&
\end{flalign}
\begin{flalign}
\mathcal{I}_h=\left[ g^A\cdot \Gamma^{L}\cdot g^R
\right]_{n',s';n,s}^{\varepsilon-\omega_1}\left[D^{R}-D^{A}\right]_{p',n';n,p}^{\omega_1}
\left[g^{R}\cdot\Gamma^{R}\cdot S \cdot g^{A}\right]_{p,s;p',s'}^{\varepsilon};
&&
\end{flalign}
\begin{flalign}
\mathcal{I}_i=&\left[g^R\cdot \Gamma^{R}\cdot  S \cdot g^A\cdot \Gamma^{L}\cdot g^R-
g^A\cdot \Gamma^{L}\cdot g^R\cdot \Gamma^{R}\cdot S \cdot g^A
\right]_{n',s';n,s}^{\varepsilon}\left[D^{R}-D^{A}\right]_{n,p;p',n'}^{\omega_1}\\\nonumber
&\hspace{10mm}\times
G_{p,s,m,\sigma}^{R}(\varepsilon-\omega_1)
\left[D^{R}-D^{A}\right]_{m,q;q',m'}^{\omega_2}
\left[g^{R}\cdot\Gamma^{L}\cdot g^{A}\right]_{q,\sigma;q',\sigma'}^{\varepsilon-\omega_1-\omega_2}
G_{m',\sigma';p',s'}^{A}(\varepsilon-\omega_1);
&&
\end{flalign}
\begin{flalign}
\mathcal{I}_j=&\left[g^R\cdot \Gamma^{R}\cdot S \cdot g^A\cdot \Gamma^{L}\cdot g^R-
g^A\cdot \Gamma^{L}\cdot g^R\cdot \Gamma^{R}\cdot S \cdot g^A
\right]_{n',s';n,s}^{\varepsilon-\omega_1-\omega_2}\left[D^{R}-D^{A}\right]_{p',n';n,p}^{\omega_2}\\\nonumber
&\hspace{10mm}\times
G_{p,s;m,\sigma}^{R}(\varepsilon-\omega_1)
\left[D^{R}-D^{A}\right]_{q',m';m,q}^{\omega_1}
\left[g^{R}\cdot\Gamma^{R}\cdot g^{A}\right]_{q,\sigma;q',\sigma'}^{\varepsilon}
G_{m',\sigma';p',s'}^{A}(\varepsilon-\omega_1);
&&
\end{flalign}
\begin{flalign}
\mathcal{I}_k=& \left[g^A\cdot \Gamma^{L}\cdot g^R
\right]_{n',s';n,s}^{\varepsilon-\omega_1-\omega_2}\left[D^{R}-D^{A}\right]_{p',n';n,p}^{\omega_2}
\\\nonumber
&\hspace{10mm}\times
G_{p,s;m,\sigma}^{R}(\varepsilon-\omega_1)
\left[D^{R}-D^{A}\right]_{q',m';m,q}^{\omega_1}
\left[g^{R}\cdot\Gamma^{R}\cdot S \cdot g^{A}\right]_{q,\sigma;q',\sigma'}^{\varepsilon}
G_{m',\sigma';p',s'}^{A}(\varepsilon-\omega_1).
&&
\end{flalign}
\end{subequations}
\end{widetext}
The matrix $S_{n,s;n',s'}=\delta_{n,R}\delta_{n',n}\delta_{s,s'}\delta_{s,s_0}$ polarizes the incoming current at the right lead (at $n=Len$) and the current is measured at left lead ($n=1$). For clarity, we write $\mathcal{I}_{\alpha}$ with the same energy dependence as the diagrams used for expressing the current in Ref.~\onlinecite{Klein}.  

\section{Necessity of non-equilibrium for AMR} \label{App:non-eq}

In the main text, we state that the condition 
\begin{align}
	G_{x,s;x',s'}^{R,A}(B)\neq G_{x',-s';x,-s}^{R,A}(-B)
\end{align}
must hold in order to observe AMR in the two terminal experimental setups involving a single magnetic lead. Within perturbation theory, we can show that the difference between the two retarded GFs is proportional to
\begin{align}\nonumber 
&G^R_{x,s;x',s'} \left( B \right) - G^R_{x',-s';x,-s} \left( -B \right) \hspace{-0.5mm}\propto \hspace{-0.5mm} \int\frac{d\omega}{4\pi}\left[f_{\mathrm{L}}^{\varepsilon-\omega} - f_{\mathrm{R}}^{\varepsilon-\omega}\right] \\
& D^R_{x_{1},x_{2};x_{3},x_{4}}(\omega)g^R_{x,s;x_{1},s_{1}}(\varepsilon,B)g^R_{x_4,s_2;x',s'}(\varepsilon,B)\\\nonumber
&\times\left[ g^R \cdot(\Gamma_L-\Gamma_R) \cdot g^A - g^A \cdot(\Gamma_L-\Gamma_R)\cdot g^R \right]_{x_{2},s_{1};x_{3},s_{2}}^{\epsilon-\omega,B} 
\end{align}
Here one can see explicitly that the difference depends on non-equilibrium conditions, in addition to interactions. In the derivation we used the identity $G^A-G^R=G^<-G^>$ and the relation $g_{x,s;x',s'}^{R,A}(B) = g_{x',-s';x,-s}^{R,A}(-B)$ valid for non-interacting systems. Similar results hold for the advanced GFs.

\begin{small}

\end{small}

\end{document}